\documentclass[12pt,dvipsnames]{article}
 
\usepackage{iftex}
\ifLuaTeX
  \usepackage{polyglossia} 
  \usepackage{fontspec} 
  \setmainlanguage{english}
  \usepackage[nosort]{cite}
\else 
  \usepackage[utf8]{inputenc}
  \usepackage[T1]{fontenc}
  \usepackage[nosort]{cite}
  \usepackage[english]{babel} 
\fi

\usepackage[top=2cm,bottom=2cm,outer=2.5cm,inner=2.5cm,marginparwidth=2.5cm]{geometry}
\usepackage{fvextra} 
\usepackage{authblk} 

\ifLuaTeX
  \usepackage[unicode]{hyperref}
  \usepackage{pdftexcmds} 
  \makeatletter
  \let\pdfstrcmp\pdf@strcmp
  \makeatother
\else
  \usepackage{hyperref}
\fi

\usepackage{amsmath}
\usepackage{amsfonts}
\usepackage{amssymb}
\usepackage{empheq} 
\usepackage{romanbar} 
\usepackage{dsfont} 

\usepackage{lmodern} 
\usepackage{caption}
\usepackage{subcaption} 
\usepackage{titlesec} 
\usepackage{fancyhdr} 
\usepackage{abstract} 
\usepackage[super]{nth}
\usepackage{tocloft} 

\usepackage{graphicx} 
\usepackage{multirow} 
\usepackage{array} 
\usepackage{tikz}
\usepackage[compat=1.1.0]{tikz-feynman} 
\usepackage{pgf}
\usepackage{pgfplots} 
\usepackage{booktabs}

\usetikzlibrary{arrows,shapes,calc,intersections,fadings,positioning,shadows.blur,decorations.pathreplacing,external,backgrounds}
\usepackage{currfile} 
\tikzset{cross/.style={cross out, draw, 
         minimum size=2*(#1-\pgflinewidth), 
         inner sep=0pt, outer sep=0pt}}

\immediate\write18{mkdir -p TikZ}     
\tikzexternaldisable
\newcommand*\setmyname{
  \expandafter\tikzsetfigurename\expandafter{\currfilebase-}%
}
\AtBeginEnvironment{tikzpicture}{\setmyname}

\hypersetup{
pdfstartview={FitH},
pdftitle={},
pdfauthor={},
pdfsubject={},
pdfcreator={},
pdfkeywords={},
colorlinks=true,
citecolor=MidnightBlue,
linkcolor=MidnightBlue,
urlcolor=MidnightBlue
}

\numberwithin{equation}{section}
\newcommand\frontmatter{%
  \clearpage
  \pagenumbering{roman}
}

\newcommand\mainmatter{%
  \clearpage
  \pagenumbering{arabic}
}

\usepackage{mathtools}

\DeclareMathOperator{\tr}{tr}
\newcommand{\vev}[1]{\left\langle #1 \right\rangle}

\DeclareMathOperator{\vol}{\mathrm{vol}}

\newcommand{\RicciScalar}{\mathcal{R}}

\newcommand{\IIFundForm}{\Romanbar{2}}
\newcommand{\diff}{\mathrm{d}}
\newcommand{\fin}{\text{finite}}

\def\bD {\mathbb{D}}

\def\bO {\mathbb{O}}

\def\bR {\mathbb{R}}

\def\cN{{\mathcal{N}}}

\def\cO{{\mathcal{O}}}

\newcommand{\bea}{\begin{eqnarray}}
\newcommand{\eea}{\end{eqnarray}}
\newcommand{\beq}{\begin{equation}}
\newcommand{\eeq}{\end{equation}}
\newcommand{\bal}{\begin{equation}\begin{aligned}}
\newcommand{\eal}{\end{aligned} \end{equation}}

\title{Surface defects in the $O(N)$ model}
\author{Maxime
  Tr\'epanier\thanks{\href{mailto:trepanier.maxime@gmail.com}{trepanier.maxime@gmail.com}}}
\affil{\it Department of Mathematics, King's College London,\\London, WC2R
2LS, United Kingdom}

\date{}

\tikzset{cross/.style={path picture={
  \draw[black]
        (path picture bounding box.south east) --
        (path picture bounding box.north west)
        (path picture bounding box.south west) --
        (path picture bounding box.north east);}}}

\newcommand{\FDfa}{%
\begin{tikzpicture}[scale=0.7,baseline=(current bounding box.center)]
    \draw[thick] (0,0) -- (2,0);
    \node[circle,fill=black,inner sep=0.1em] at (1,0) (p1) {};
    \draw[dashed] ($(p1)+(-0.03,0)$) -- ($(p1)+(-0.03,1.5)$);
    \draw[dashed] ($(p1)+(0.03,0)$) -- ($(p1)+(0.03,1.5)$);
  \end{tikzpicture}
}
\newcommand{\FDfb}{%
\begin{tikzpicture}[scale=0.7,baseline=(current bounding box.center)]
    \draw[thick] (0.5,0) -- (2.5,0);
    \node[circle,fill=black,inner sep=0.1em] at (1,0) (p1) {};
    \node[circle,fill=black,inner sep=0.1em] at (2,0) (p2) {};
    \path[dashed,in=270,out=90] ($(p1)+(-0.03,0)$) edge (1.47,1.5);
    \path[dashed,in=270,out=90] ($(p2)+(0.03,0)$) edge (1.53,1.5);
    \path[dashed, bend left] ($(p1)+(0.03,0)$) edge ($(p2)-(0.03,0)$);
  \end{tikzpicture}
}
\newcommand{\FDfba}{%
\begin{tikzpicture}[scale=0.7,baseline=(current bounding box.center)]
    \draw[thick] (0,0) -- (3,0);
    \node[circle,fill=black,inner sep=0.1em] at (0.75,0) (p1) {};
    \node[circle,fill=black,inner sep=0.1em] at (1.5,0) (p2) {};
    \node[circle,fill=black,inner sep=0.1em] at (2.25,0) (p3) {};
    \path[dashed,in=270,out=90] ($(p1)+(-0.03,0)$) edge (1.47,1.5);
    \path[dashed,in=270,out=90] ($(p3)+(0.03,0)$) edge (1.53,1.5);
    \path[dashed, bend left] ($(p1)+(0.03,0)$) edge ($(p2)-(0.03,0)$);
    \path[dashed, bend left] ($(p2)+(0.03,0)$) edge ($(p3)-(0.03,0)$);
  \end{tikzpicture}
}
\newcommand{\FDfbb}{%
\begin{tikzpicture}[scale=0.7,baseline=(current bounding box.center)]
    \draw[thick] (0.5,0) -- (2.5,0);
    \node[circle,fill=black,inner sep=0.1em] at (1,0) (p1) {};
    \node[circle,fill=black,inner sep=0.1em] at (2,0) (p2) {};
    \path[dashed,in=0,out=90] ($(p1)+(-0.03,0)$) edge (0.5,0.2);
    \path[dashed,in=180,out=90] ($(p2)+(0.03,0)$) edge (2.5,0.2);
    \path[dashed, bend left] ($(p1)+(0.03,0)$) edge ($(p2)-(0.03,0)$);
    \node at (1.5,0.3) {\scriptsize $m$};
  \end{tikzpicture}
}
\newcommand{\FDfbc}{%
\begin{tikzpicture}[scale=0.7,baseline=(current bounding box.center)]
    \draw[thick] (0,0) -- (3,0);
    \node[circle,fill=black,inner sep=0.1em] at (0.75,0) (p1) {};
    \node[circle,fill=black,inner sep=0.1em] at (1.5,0) (p2) {};
    \node[circle,fill=black,inner sep=0.1em] at (2.25,0) (p3) {};
    \path[dashed,in=0,out=90] ($(p1)+(-0.03,0)$) edge (0.25,0.2);
    \path[dashed,in=180,out=90] ($(p3)+(0.03,0)$) edge (2.75,0.2);
    \path[dashed, bend left] ($(p1)+(0.03,0)$) edge ($(p2)-(0.03,0)$);
    \path[dashed, bend left] ($(p2)+(0.03,0)$) edge ($(p3)-(0.03,0)$);
    \node at (1.16,0.3) {\scriptsize $1$};
    \node at (1.87,0.3) {\scriptsize $m$};
  \end{tikzpicture}
}
\newcommand{\FDfbd}{%
\begin{tikzpicture}[scale=0.7,baseline=(current bounding box.center)]
    \draw[thick] (0.5,0) -- (2.5,0);
    \node[circle,fill=black,inner sep=0.1em] at (1,0) (p1) {};
    \node[circle,fill=black,inner sep=0.1em] at (2,0) (p2) {};
    \path[dashed,in=0,out=90] ($(p1)+(-0.03,0)$) edge (0.5,0.2);
    \path[dashed,in=180,out=90] ($(p2)+(0.03,0)$) edge (2.5,0.2);
    \path[dashed, bend left] ($(p1)+(0.03,0)$) edge ($(p2)-(0.03,0)$);
    \node at (1.5,0.3) {\scriptsize $m+1$};
  \end{tikzpicture}
}
\newcommand{\FDfc}{%
  \begin{tikzpicture}[scale=0.7,baseline=(current bounding box.center)]
    \draw[thick] (0,0) -- (2,0);
    \node[circle,fill=black,inner sep=0.1em] at (1,0) (p1) {};
    \node[circle,fill=black,inner sep=0pt,minimum size=2pt] at (1,1) (p2) {};
    \path[dashed,in=225,out=135] ($(p1)+(-0.03,0)$) edge ($(p2)+(-0.03,0)$);
    \path[dashed,in=315,out=45] ($(p1)+(0.03,0)$) edge ($(p2)+(0.03,0)$);
    \path[dashed,out=135,in=270] ($(p2)+(-0.03,0)$) edge (0.97,1.5);
    \path[dashed,out=45,in=270] ($(p2)+(0.03,0)$) edge (1.03,1.5);
  \end{tikzpicture}
}
\newcommand{\FDfca}{%
  \begin{tikzpicture}[scale=0.7,baseline=(current bounding box.center)]
    \draw[thick] (0,0) -- (2,0);
    \coordinate (p1) at (1,0);
    \draw[dashed] ($(p1)+(-0.03,0)$) -- ($(p1)+(-0.03,1.5)$);
    \draw[dashed] ($(p1)+(0.03,0)$) -- ($(p1)+(0.03,1.5)$);
    \draw (p1) circle (4pt);
    \node[draw,circle,black,cross,fill=white,inner sep=0pt,minimum size=6pt] at (1,0) (p1) {};
  \end{tikzpicture}
}
\newcommand{\FDfcb}{%
  \begin{tikzpicture}[scale=0.7,baseline=(current bounding box.center)]
    \draw[thick] (0,0) -- (2,0);
    \coordinate (p1) at (1,0.75) {};
    \draw[dashed] ($(0.97,0)$) -- ($(0.97,1.5)$);
    \draw[dashed] ($(1.03,0)$) -- ($(1.03,1.5)$);
    \node[draw,circle,black,cross,fill=white,inner sep=0pt,minimum size=6pt] at (p1) {};
  \end{tikzpicture}
}

\newcommand{\FDsa}{%
\begin{tikzpicture}[scale=0.7,baseline=(current bounding box.center)]
  \draw[thick] (0,0) -- (2,0);
  \node[circle,fill=black,inner sep=0.1em] at (1,0) (p1) {};
  \draw (p1) -- (1,1.5);
\end{tikzpicture}
}
\newcommand{\FDsb}{%
\begin{tikzpicture}[scale=0.7,baseline=(current bounding box.center)]
  \draw[thick] (0.5,0) -- (2.5,0);
  \node[circle,fill=black,inner sep=0.1em] at (1,0) (p1) {};
  \node[circle,fill=black,inner sep=0.1em] at (2,0) (p2) {};
  \draw (p1) -- (1.5,0.5) -- (p2);
  \draw (1.5,0.5) -- (1.5,1.5);
\end{tikzpicture}
}
\newcommand{\FDsc}{%
\begin{tikzpicture}[scale=0.7,baseline=(current bounding box.center)]
  \draw[thick] (0,0) -- (2,0);
  \node[circle,fill=black,inner sep=0.1em] at (1,0) (p1) {};
  \draw (p1) -- (1,0.5);
  \draw (1,0.5) arc[radius=0.25,start angle=-90, end angle=270];
  \draw (1,1) -- (1,1.5);
\end{tikzpicture}
}
\newcommand{\FDsd}{%
\begin{tikzpicture}[scale=0.7,baseline=(current bounding box.center)]
  \draw[thick] (0,0) -- (2,0);
  \node[circle,fill=black,inner sep=0.1em] at (1,0) (p1) {};
  \draw (p1) -- (1,0.5);
  \draw[dashed] (1,0.5) arc[radius=0.25,start angle=-90, end angle=270];
  \draw (1,1) -- (1,1.5);
\end{tikzpicture}
}
\newcommand{\FDse}{%
\begin{tikzpicture}[scale=0.7,baseline=(current bounding box.center)]
  \draw[thick] (0.25,0) -- (2.75,0);
  \node[circle,fill=black,inner sep=0.1em] at (0.75,0) (p1) {};
  \node[circle,fill=black,inner sep=0.1em] at (1.5,0) (p2) {};
  \node[circle,fill=black,inner sep=0.1em] at (2.25,0) (p3) {};
  \node[inner sep=0pt,outer sep=0pt,minimum size=0pt] at (1.87,0.5) (p5) {};
  \node[inner sep=0pt,outer sep=0pt,minimum size=0pt] at (1.5,1) (p6) {};
  \draw (p1) -- (p6) -- (1.5,1.5);
  \draw (p2) -- (p5) -- (p3);
  \draw (p5) -- (p6);
\end{tikzpicture}
}
\newcommand{\FDsf}{%
\begin{tikzpicture}[scale=0.7,baseline=(current bounding box.center)]
  \draw[thick] (0.5,0) -- (2.5,0);
  \node[circle,fill=black,inner sep=0.1em] at (1,0) (p1) {};
  \node[circle,fill=black,inner sep=0.1em] at (2,0) (p2) {};
  \node[inner sep=0pt] at (1.5,0.5) (p3) {};
  \draw[dashed] (p1) -- (p3) -- (p2);
  \draw (p3) -- (1.5,1.5);
\end{tikzpicture}
}
\newcommand{\FDsg}{%
\begin{tikzpicture}[scale=0.7,baseline=(current bounding box.center)]
  \draw[thick] (0.25,0) -- (2.75,0);
  \node[circle,fill=black,inner sep=0.1em] at (0.75,0) (p1) {};
  \node[circle,fill=black,inner sep=0.1em] at (1.5,0) (p2) {};
  \node[circle,fill=black,inner sep=0.1em] at (2.25,0) (p3) {};
  \node[inner sep=0pt,outer sep=0pt,minimum size=0pt] at (1.87,0.5) (p5) {};
  \node[inner sep=0pt,outer sep=0pt,minimum size=0pt] at (1.5,1) (p6) {};
  \draw (p1) -- (p6) -- (1.5,1.5);
  \draw[dashed] (p2) -- (p5) -- (p3);
  \draw (p5) -- (p6);
\end{tikzpicture}
}
\newcommand{\FDsh}{%
\begin{tikzpicture}[scale=0.7,baseline=(current bounding box.center)]
  \draw[thick] (0.25,0) -- (2.75,0);
  \node[circle,fill=black,inner sep=0.1em] at (0.75,0) (p1) {};
  \node[circle,fill=black,inner sep=0.1em] at (1.5,0) (p2) {};
  \node[circle,fill=black,inner sep=0.1em] at (2.25,0) (p3) {};
  \node[inner sep=0pt,outer sep=0pt,minimum size=0pt] at (1.87,0.5) (p5) {};
  \node[inner sep=0pt,outer sep=0pt,minimum size=0pt] at (1.5,1) (p6) {};
  \draw[dashed] (p1) -- (p6) -- (p5) -- (p2);
  \draw (p3) -- (p5);
  \draw (p6) -- (1.5,1.5);
\end{tikzpicture}
}

\newcommand{\FDsca}{%
\begin{tikzpicture}[scale=0.7,baseline=(current bounding box.center)]
  \draw[thick] (0,0) -- (2,0);
  \coordinate (p1) at (1,0);
  \draw (p1) -- (1,1.5);
  \node[draw,circle,black,cross,fill=white,inner sep=0pt,minimum size=6pt] at (p1) {};
\end{tikzpicture}
}
\newcommand{\FDscb}{%
\begin{tikzpicture}[scale=0.7,baseline=(current bounding box.center)]
  \draw[thick] (0,0) -- (2,0);
  \node[circle,fill=black,inner sep=0.1em] at (1,0) (p1) {};
  \draw (p1) -- (1,1.5);
  \node[draw,circle,black,cross,fill=white,inner sep=0pt,minimum size=6pt] at
  (1,0.75) (p2) {};
\end{tikzpicture}
}
\newcommand{\FDscc}{%
\begin{tikzpicture}[scale=0.7,baseline=(current bounding box.center)]
  \draw[thick] (0.5,0) -- (2.5,0);
  \node[circle,fill=black,inner sep=0.1em] at (1,0) (p1) {};
  \coordinate (p2) at (2,0);
  \draw (p1) -- (1.5,0.5) -- (p2);
  \draw (1.5,0.5) -- (1.5,1.5);
  \node[draw,circle,black,cross,fill=white,inner sep=0pt,minimum size=6pt] at
  (p2) {};
\end{tikzpicture}
}

\newcommand{\FDsj}{%
\begin{tikzpicture}[scale=0.7,baseline=(current bounding box.center)]
  \draw[thick] (0,0) -- (2,0);
  \node[circle,fill=black,inner sep=0.1em] at (1,0) (p1) {};
  \draw[dashed] (p1) -- ($(p1)+(0,1.5)$);
\end{tikzpicture}
}
\newcommand{\FDsk}{%
\begin{tikzpicture}[scale=0.7,baseline=(current bounding box.center)]
  \draw[thick] (0.5,0) -- (2.5,0);
  \node[circle,fill=black,inner sep=0.1em] at (1,0) (p1) {};
  \node[circle,fill=black,inner sep=0.1em] at (2,0) (p2) {};
  \draw[dashed] (p1) -- (1.5,0.5) -- (1.5,1.5);
  \draw (p2) -- (1.5,0.5);
\end{tikzpicture}
}
\newcommand{\FDsl}{%
\begin{tikzpicture}[scale=0.7,baseline=(current bounding box.center)]
  \draw[thick] (0.25,0) -- (2.75,0);
  \node[circle,fill=black,inner sep=0.1em] at (0.75,0) (p1) {};
  \node[circle,fill=black,inner sep=0.1em] at (1.5,0) (p2) {};
  \node[circle,fill=black,inner sep=0.1em] at (2.25,0) (p3) {};
  \node[inner sep=0pt,outer sep=0pt,minimum size=0pt] at (1.87,0.5) (p5) {};
  \node[inner sep=0pt,outer sep=0pt,minimum size=0pt] at (1.5,1) (p6) {};
  \draw[dashed] (p1) -- (p6) -- (1.5,1.5);
  \draw[dashed] (p2) -- (p5) -- (p3);
  \draw (p5) -- (p6);
\end{tikzpicture}
}
\newcommand{\FDsm}{%
\begin{tikzpicture}[scale=0.7,baseline=(current bounding box.center)]
  \draw[thick] (0.25,0) -- (2.75,0);
  \node[circle,fill=black,inner sep=0.1em] at (0.75,0) (p1) {};
  \node[circle,fill=black,inner sep=0.1em] at (1.5,0) (p2) {};
  \node[circle,fill=black,inner sep=0.1em] at (2.25,0) (p3) {};
  \node[inner sep=0pt,outer sep=0pt,minimum size=0pt] at (1.87,0.5) (p5) {};
  \node[inner sep=0pt,outer sep=0pt,minimum size=0pt] at (1.5,1) (p6) {};
  \draw[dashed] (p1) -- (p6) -- (1.5,1.5);
  \draw (p2) -- (p5) -- (p3);
  \draw (p5) -- (p6);
\end{tikzpicture}
}
\newcommand{\FDsn}{%
\begin{tikzpicture}[scale=0.7,baseline=(current bounding box.center)]
  \draw[thick] (0.25,0) -- (2.75,0);
  \node[circle,fill=black,inner sep=0.1em] at (0.75,0) (p1) {};
  \node[circle,fill=black,inner sep=0.1em] at (1.5,0) (p2) {};
  \node[circle,fill=black,inner sep=0.1em] at (2.25,0) (p3) {};
  \node[inner sep=0pt,outer sep=0pt,minimum size=0pt] at (1.87,0.5) (p5) {};
  \node[inner sep=0pt,outer sep=0pt,minimum size=0pt] at (1.5,1) (p6) {};
  \draw[dashed] (p2) -- (p5) -- (p6) -- (1.5,1.5);
  \draw (p1) -- (p6);
  \draw (p3) -- (p5);
\end{tikzpicture}
}
\newcommand{\FDsp}{%
\begin{tikzpicture}[scale=0.7,baseline=(current bounding box.center)]
  \draw[thick] (0,0) -- (2,0);
  \node[circle,fill=black,inner sep=0.1em] at (1,0) (p1) {};
  \draw[dashed] (p1) -- (1,0.5);
  \draw[dashed] (1,0.5) arc[radius=0.25,start angle=-90, end angle=90];
  \draw (1,0.5) arc[radius=0.25,start angle=270, end angle=90];
  \draw[dashed] (1,1) -- (1,1.5);
\end{tikzpicture}
}

\newcommand{\FDfi}{%
\begin{tikzpicture}[scale=0.7,baseline=(current bounding box.center)]
    \draw[thick] (-1,-0.5) -- (1,-0.5);
    \draw (-0.5,-0.5) -- (-0.5,0);
    \draw (0,-0.5) -- (0,0);
    \draw (0.5,-0.5) -- (0.5,0);
    \draw (0,0.5) -- (0,1);
    \draw[dashed] (-0.5,0) -- (0,0) -- (0.5,0) -- (0,0.5) -- (-0.5,0);
  \end{tikzpicture}
}

\newcommand{\FDprops}{%
\begin{tikzpicture}[scale=0.7,baseline={([yshift=-3pt]current bounding
  box.center)}]
    \draw[thick] (0,0) -- (2,0);
  \end{tikzpicture}
}
\newcommand{\FDpropphia}{%
  \begin{tikzpicture}[scale=0.7,baseline={([yshift=-3pt]current bounding
    box.center)}]
    \path[dashed,in=155,out=35] (0,0.03) edge (2,0.03);
    \path[dashed,in=215,out=-35] (0,-0.03) edge (2,-0.03);
    \node[circle,fill=black,inner sep=0.1em] at (0,0) (p1) {};
    \node[circle,fill=black,inner sep=0.1em] at (2,0) (p1) {};
  \end{tikzpicture}
}
\newcommand{\FDpropphib}{%
\begin{tikzpicture}[scale=0.7,baseline={([yshift=-3pt]current bounding
  box.center)}]
    \path[dashed,in=155,out=35] (0,0.03) edge (1,0.03);
    \path[dashed,in=155,out=35] (1,0.03) edge (2,0.03);
    \path[dashed,out=-35,in=215] (0,-0.03) edge (1,-0.03);
    \path[dashed,out=-35,in=215] (1,-0.03) edge (2,-0.03);
    \node[circle,fill=black,inner sep=0.1em] at (0,0) {};
    \node[circle,fill=black,inner sep=0.1em] at (1,0) {};
    \node[circle,fill=black,inner sep=0.1em] at (2,0) {};
  \end{tikzpicture}
}

\begin{document}

\frontmatter
\maketitle
\thispagestyle{empty}

\begin{abstract}
  I study the two-dimensional defects of the $d$ dimensional critical $O(N)$
  model and the defect RG flows between them. By combining the
  $\epsilon$-expansion around $d = 4$ and $d = 6$ as well as large $N$
  techniques, I find new conformal defects and examine their behavior across
  dimensions and at various $N$. I discuss how some of these fixed points relate
  to the known ordinary, special and extraordinary transitions in the 3d theory,
  as well as examine the presence of new symmetry breaking fixed points
  preserving an $O(p) \times O(N-p)$ subgroup of $O(N)$ for $N \le N_c$ (with
  the estimate $N_c = 6$). I characterise these fixed points by obtaining their
  conformal anomaly coefficients, their 1-point functions and comment on the
  calculation of their string potential.  These results establish surface
  operators as a viable approach to the characterisation of interface critical
  phenomena in the 3d critical $O(N)$ model.
\end{abstract}

\mainmatter

\section{Introduction and summary}

The problem of classifying boundaries and interfaces of the 3d critical
$O(N)$ model has attracted considerable attention since the pioneering works
of~\cite{AJBray_1977,ohno,cardy1984conformal,McAvity:1995zd}. The critical
exponents characterising some of the possible boundary phases were obtained early on
through a combination of field theory techniques and found experimental
verification in phenomena such as the critical adsorption of fluids on
walls~\cite{fisher1978wall,liu1989universal,diehl1994critical,law1994surface,floter1995universal},
see~\cite{Diehl:1996kd} for an early review. Yet, still today the
characterisation of the various phases and their organisation into a phase
diagram for general $N$ is not settled~\cite{Metlitski:2020cqy,Krishnan:2023cff}.

Although the $d=3$ critical $O(N)$ model is the most interesting for describing
physical systems, it is not directly accessible to perturbative methods and the
theory is often studied by analytically continuing $d$ to the range $2 < d < 6$.
In this context it is an interesting question to understand how to describe
boundaries and interfaces away from $d=3$.
The most established method to do so is to study boundaries/interfaces in $d$
dimensions, which corresponds to keeping the codimension of the defect fixed.
For boundaries this is dictated by the topology of spacetime and that approach
has been the subject of numerous works such
as~\cite{ohno,McAvity:1995zd,PhysRevB.25.3283,Gliozzi:2015qsa,Dey:2020lwp,Nishioka:2022odm}.
For interfaces, however, there is another natural choice: we can keep the
dimension of the defect fixed and study surface defects in the $d$ dimensional
$O(N)$ model.%
\footnote{
  This applies to ``topologically trivial'' interfaces, where we have a copy of
  the $O(N)$ model of each side of the defect.
}
This is the focus of this paper.

The origin of these surface defects is easy to explain. In any CFT, we may consider a
set of local operators $\cO_I$ with conformal dimension $\Delta_{\cO_I}$ and
construct a defect by integrating local operators over a plane with some
coupling constants $u^I$ to be fixed shortly
\begin{align}
  D
  =
  \exp \left[ - \int_{\bR^2} \diff^2 \tau u^I \cO_I \right]\,.
  \label{eqn:defectCPT}
\end{align}
Unless the operators have $\Delta_{\cO_I} = 2$ exactly, the
couplings $u^I$ get renormalised and give rise to a defect RG flow (dRG flow).
The beta function that governs their renormalisation near the trivial defect
$u^I = 0$ is well understood and given by~\cite{Cardy:1988cwa,Klebanov:2011gs}
\begin{align}
  \beta_{u^I}
  =
  (\Delta_{\cO_I} - 2) u^I
  + \pi {C^I}_{JK} u^J u^K
  + \dots
  \label{eqn:defectbetafunction}
\end{align}
where $C_{IJK}$ are the structure constants of the bulk theory for $\cO_I$ and the indices
are raised by the Zamolodchikov metric for $\cO_I$.
The nontrivial zeros $u^I_*$ of this beta function then correspond to nontrivial
conformal surface defects.

In this paper, I undertake the exploration of surface defects and their dRG
flows in the $O(N)$ model through perturbative methods. By combining the
$\epsilon$-expansion as well as exploiting large $N$ techniques, I uncover a
rich structure of dRG flows and fixed points corresponding to new defects, as
well as phenomena such as the appearance/annihilation of fixed points as we vary
the parameters $d$ and $N$.

The simplest setting in which one can study surface defects is perhaps at
$d = 4 - \epsilon$ using the $\epsilon$-expansion, and this is presented in
section~\ref{sec:pert4d}. There, surface defects
naturally arise from integrating the operator $\varphi^i \varphi^j$ over a plane
as in~\eqref{eqn:defectCPT} ($i = 1,\cdots,N$). In addition to the $O(N)$ symmetric defect
$D_N$ coupling to $\varphi^k \varphi^k$, I find fixed points corresponding to
conformal defects $D_p$ preserving a subgroup $O(p) \times O(N-p)$ of the full
$O(N)$ symmetry (with $0 \le p \le N$).
These defects have an analog in the free $O(N)$ model analysed
in~\cite{Shachar:2022fqk}, where one can engineer a conformal defect preserving
the same symmetry by taking $p$ scalars at one zero of the beta function ($u_* = \pi
\epsilon$) and the rest at the other ($u_* = 0$). As for the free theory, these defects can
be thought of as saddle points of a symmetry breaking dRG flow, with the stable
fixed point the symmetry preserving defect $D_N$. Surprinsingly however, here
the symmetry breaking fixed points only exist for small enough $N \le N_c$, with
$N_c = 6$, to first order in $\epsilon$.

Away from $d = 4$ these defects can be reliably studied using the
large $N$ expansion. In section~\ref{sec:largen} I examine the behavior of the
stable fixed point $D_N$ across dimensions. This is particularly simple when
studying the theory through the Hubbard-Stratonovich transformation, as the 
defect couples directly to the Hubbard-Stratonovich field $\sigma$. Taking $d
\to 3$ leads to a divergence in the defect coupling $u_*$ (the zero
of the beta function of the coupling to $\sigma$), which signals a
change of scaling with respect to $N$. I argue that $D_N$ becomes the fixed
point of the well-known ordinary transition in $d = 3$, while the trivial defect
is known to be the special transition~\cite{AJBray_1977}.

In addition to the ordinary and special transitions, the 3d $O(N)$ model has the
extraordinary (or normal) transition, which is characterised by its breaking of
$O(N)$ symmetry to $O(N-1)$. A setup with perturbative control to study
this defect is the $\epsilon$-expansion around $d = 6$, where the breaking is
naturally realised by coupling the defect to a fundamental scalar $\varphi^1$.
This is the analog of the pinning defect studied in the context of line
operators~\cite{Cuomo:2021kfm}. We find that near $d = 6$ the fixed point
corresponds to a non-unitary surface defect, at least at large $N$. It is
interesting to note that the same is true of the boundary defect describing the
extraordinary fixed point around $d = 6$~\cite{Giombi:2020rmc}.

It is rather encouraging to recover the known classes of interfaces at
$d = 3$ directly from surface defects: it suggests that surface defects are a
viable alternative to describing interface critical phenomena in $d = 3$.
Now, studying surface operators instead of boundaries has certain advantages.
The main property of surface defects (and indeed interfaces in $d=3$) is their
conformal anomaly governing the UV divergences of their expectation value, here
this is manifest for any $d$. For a surface defect $D_\Sigma$ defined over a
surface $\Sigma$ inside a $d$-dimensional manifold of metric $G$, the
expectation value receives a universal contribution of the
form~\cite{Schwimmer:2008yh}
\begin{align}
  \log\vev{D_\Sigma}
  =
  \frac{\log \tilde\epsilon}{4 \pi} \int \vol_\Sigma 
  \left[ a \RicciScalar^\Sigma + b_1 \tr \tilde{\IIFundForm}^2 + b_2
  \tr{W}
  + c (\partial n)^2 \right]
  + \fin,
  \label{eqn:anomalycoeff}
\end{align}
where $\tilde{\varepsilon}$ is a UV cutoff, $\RicciScalar^\Sigma$ is the 2d
Ricci scalar on $\Sigma$, $\tilde\IIFundForm$ is the traceless part of the
second fundamental form squared, $\tr W$ is the trace of the pullback of the
Weyl tensor on $\Sigma$, and $(\partial n)^2$ is a term relevant for symmetry
breaking surfaces; its precise form is given in section~\ref{sec:anomaly}. Note
that in $d=3$ the Weyl tensor vanishes identically and that term is absent. A
review of the geometric invariants and the various basis used for conformal
anomalies can be found in~\cite{Drukker:2020dcz} (see
also~\cite{Herzog:2020wlo}).

The anomaly coefficients $a, b_1, b_2, c$ do not depend on the
geometry, but may depend on $d, N$.  They are particularly interesting because
they are constrained by unitarity and appear in a wide variety of observables.
Perhaps most importantly, the coefficient $a$ is known to obeys an
$a$-theorem~\cite{Jensen:2015swa} (see also~\cite{Wang:2020xkc,Shachar:2022fqk})
\begin{align}
  a_{UV} > a_{IR}\,,
  \label{eqn:atheorem}
\end{align}
which provides an important diagnosis of dRG flows.
The perturbative dRG flows presented below provide new nontrivial examples where
the $a$-theorem applies and is indeed satisfied. More interesting are the cases
where the perturbative analysis falls short---still we expect that the IR fixed
point corresponds to the defect with lowest value of $a$ allowed by symmetry.

The coefficients $b_1, b_2, c$ are also constrained by unitarity to have
definite signs. $b_1$ and $c$ are respectively associated to the displacement
operator $\bD$ (tilt operator $\bO$) arising from the breaking of translation
symmetry (resp. the $O(N)$ symmetry).
Concretely, the displacement operator appears as a contact term for the
conservation law of the stress tensor along a direction orthogonal to the defect
$x_\perp^m$ (I write $D[\hat{\cO}]$ to denote the insertion of the local operator
$\hat{\cO}$ on the defect $D$)
\begin{align}
  \partial_\mu T^{\mu m}(x) D
  =
  D[\bD^m(x^\parallel)] \delta^{(d-2)}(x_\perp)\,.
  \label{eqn:disp}
\end{align}
A similar equation defines the tilt operator from the $O(N)$ symmetry current
$j^\mu$. Since~\eqref{eqn:disp} fixes the normalisation of $\bD$, the
coefficient $C_\bD$ appearing in the 2-point function $\vev{D[\bD^m(\sigma)
\bD^n(0)]}$ is an observable, and is known to satisfy $- b_1 \sim C_\bD >
0$~\cite{Bianchi:2015liz}.
Similarly $c \sim C_\bO > 0$~\cite{Drukker:2020atp}.

Finally $b_2$ is related to the 1-point function of stress
tensor~\cite{Jensen:2018rxu} (see
also~\cite{Lewkowycz:2014jia,Bianchi:2015liz}), and by the ANEC is expected to
satisfy $b_2 > 0$~\cite{Jensen:2018rxu}.

I calculate the anomaly coefficients for the various defects studied here in
section~\ref{sec:anomaly}. There are other quantities associated to surface
defects, and two more observables play a role in this paper. The first one is
the 1-point function of $\varphi^2$ in the presence of the defect. The
quantity that is independent of the normalisation of $\varphi^2$ (thus can be
compared between various calculations) is $a_{\varphi^2}$
\begin{align}
  \frac{\vev{\varphi^2(x) D|_{u_*}}}{\sqrt{C_{\varphi^2}} \vev{D|_{u_*}}} =
  \frac{a_{\varphi^2}}{|x_\perp|^{\Delta_{\varphi^2}}}\,,
  \qquad
  \vev{\varphi^2(x) \varphi^2(0)}
  =
  \frac{C_{\varphi^2}}{|x|^{2\Delta_{\varphi^2}}}\,.
  \label{eqn:coeffaphi2}
\end{align}
Giving a VEV to $\varphi^2$ is akin to sourcing a mass term for $\varphi$ (here
localised at the defect), with a positive mass corresponding to $a_{\varphi^2}
\le 0$.
The agreement of $a_{\varphi^2}$ between various calculations is a nontrivial
cross-check of these results.

The second quantity of interest is the generalised string potential introduced
in~\cite{Drukker:2022beq}, which is analogous to the cusp anomalous dimension of
line operators. In particular this captures the potential density $U_0$ between a pair
of planar defects separated a distance $L$, i.e. $\vev{D|_{u_*}(L) D|_{u_*}(0)} =
\frac{U_0 \mathrm{Area}}{L^2}$.

The rest of this paper is organised as follows. Sections~\ref{sec:pert4d},
\ref{sec:largen} and~\ref{sec:pert6d} present three limits where perturbative
methods can be applied to study surface defects, and are devoted to studying
dRG flows and characterising fixed points. Section~\ref{sec:anomaly} contains the calculation of the
anomaly coefficients and the string potential.
In order to alleviate the reading of the paper, the perturbative calculations
are relegated to appendix~\ref{app:pert4d}.

\paragraph{Note added:} Shortly after this paper appeared on the arXiv, I
learned of related works to appear~\cite{Raviv-Moshe:2023yvq}
and~\cite{Giombi:2023dqs}, which overlap with parts of this paper. I am
grateful to the authors for informing me of their work and sharing a preview of
their papers with me.

\section{\texorpdfstring{$\epsilon$}{Epsilon}-expansion in $d=4-\epsilon$}
\label{sec:pert4d}

Consider the critical $O(N)$ model. In $d=4-\epsilon$ this model can be
studied in renormalised perturbation theory from the action
\begin{align}
  S =
  \int \diff^{4-\epsilon} x \left[
    \frac{1}{2} \left( \partial_\mu \varphi^i \right)^2
    + \frac{M^\epsilon \lambda}{4!} \left( \varphi^i \varphi^i \right)^2
  \right]\,,
  \label{eqn:S4d}
\end{align}
where $M$ is the renormalisation scale and we omitted the counterterms. The
theory becomes conformal when tuning the coupling constant to its critical
point
\begin{align}
  \frac{\lambda_*}{(4\pi)^2} = \frac{3}{N+8} \epsilon
  + O(\epsilon^2)\,.
  \label{eqn:lambdas}
\end{align}

We can construct a surface operator in this theory by integrating $\varphi^i
\varphi^j$ over a surface%
\footnote{More generally, one may also include the relevant operator $u^i
  \varphi^i$. We discuss such defects in $d = 6-\epsilon$ in
  section~\ref{sec:pert6d} where the dRG flows are easily accessible to
  perturbative methods.
}
\begin{align}
  D = \exp\left[ -\frac{M^\epsilon}{2} \int_{\bR^2} \diff^2 \tau
  h_{ij} \varphi^i \varphi^j \right]\,,
  \label{eqn:defnsurface4d}
\end{align}
where we included an explicit dependence on the renormalisation scale $M$ to
ensure that $h_{ij}$ is dimensionless.

The coupling $h_{ij}$ gets renormalised and obeys the beta
function~\eqref{eqn:defectbetafunction}. This is easy to see in perturbation
theory. To lowest order, there are three diagrams contributing to the
renormalisation of the coupling. They can be evaluated in dimensional
regularisation using standard methods (see appendix~\ref{app:pert4d} for
details) and read
\begin{align}
  \label{fig:renormalisationh4dl0a}
  \FDfa
  &=
  -\frac{h_{ij}}{16 \pi^3 x_\perp^2} + \dots\,,\\
  \label{fig:renormalisationh4dl0b}
  \FDfc
  &=
  \frac{\lambda}{(4\pi)^2} \frac{(h_{kk} \delta_{ij} + 2 h_{ij})}{24 \pi^3
  x_\perp^2} \left[ \frac{1}{\epsilon} +
  \frac{4 + 3 \gamma + 3 \log(\pi M^2 x_\perp^2)}{2} + \dots \right]\,,\\
  \FDfb
  &=
  \frac{h_{ik} h_{kj}}{32 \pi^4 x_\perp^2} \left[ \frac{1}{\epsilon} +
  \frac{4 + 3 \gamma + 3 \log(\pi M^2 x_\perp^2)}{2} + \dots \right]\,.
  \label{eqn:feyn4dh2}
\end{align}
Here the dashed lines are propagators for $\varphi^i$, while the solid
line represents an integral over the plane supporting the defect. $\gamma$ is
the Euler constant.

The poles in $\epsilon$ are absorbed by counterterms and ensure that the
expectation value of the composite operator $[\varphi^i \varphi^j]$ is finite,
see the corresponding diagrams in table~\ref{tab:feyn4d}.
The divergence of the second diagram is cancelled by the renormalisation of
$[\varphi^i \varphi^j]$, which in terms of irreducible representations---singlet
$S$ and symmetric traceless $T$---are given by (see
e.g.~\cite{Henriksson:2022rnm})
\begin{align}
  Z_{S} = 1 - \frac{\lambda}{(4\pi)^2} \frac{N+2}{3} \frac{1}{\epsilon} + \dots\,,
  \qquad
  Z_{T} = 1 - \frac{\lambda}{(4\pi)^2} \frac{2}{3} \frac{1}{\epsilon} + \dots
  \label{eqn:deltaphi2}
\end{align}
The remaining pole of~\eqref{eqn:feyn4dh2} can be cancelled by the counterterm
$\delta h_{ij}$
\begin{align}
  \delta h_{ij} =
  \frac{1}{\epsilon}
  \frac{h_{ik} h_{kj}}{2\pi}
  + O(h^2 \lambda, h^3)\,.
\end{align}
From these counterterms we can read the beta function for the coupling
$h_{ij}$.  Decomposing $h_{ij} = h^S \delta_{ij} + h^T_{ij}$ in irreducible
representations, the renormalised couplings $h^{S,T}$ are related to the bare
coupling $(h^{S,T})^{(0)}$ by
\begin{align}
  (h^S)^{(0)}
  =
  M^{\epsilon} \left( h^S + \delta h^S \right) Z_{S}^{-1/2}\,,
  \qquad
  (h_{ij}^T)^{(0)} = M^{\epsilon} \left( h_{ij}^T + \delta h_{ij}^T \right)
  Z_{T}^{-1/2}\,.
\end{align}
Using that bare couplings $h^{(0)}$ are independent of the renormalisation scale $M$, one
can easily obtain the beta functions for $h^S$ and $h^T$ (see e.g. the review~\cite{kleinert2001critical},
or~\cite{Cuomo:2021kfm} in the context of defects). Assembling both
representations we get
\begin{align}
  \beta_{h_{ij}}
  =
  - \epsilon h_{ij}
  + \frac{\lambda}{(4\pi)^2} \frac{h_{kk} \delta_{ij} + 2 h_{ij}}{3}
  + \frac{1}{2\pi} h_{ik} h_{kj}
  + O(h^3,h^2 \lambda, h\lambda^2) \,.
  \label{eqn:beta4d}
\end{align}

This result can be compared with~\eqref{eqn:defectbetafunction}.  The terms
linear in $h$ combine to give the conformal dimension as expected (for
reference, the dimensions of $[\varphi^i\varphi^j]$ are $\Delta_{S} = 2 -
\frac{6}{N+8} \epsilon + \dots$ and $\Delta_{T} = 2 - \frac{N+6}{N+8} \epsilon +
\dots$ respectively for the singlet and symmetric traceless
representations~\cite{Henriksson:2022rnm}). In both cases, $\Delta < 2$ so the
operator is relevant and triggers an RG flow.

The term quadratic in $h$ reproduces the structure constant of
$[\varphi^i \varphi^j]$. For this calculation we only need the leading order
contribution in $\epsilon$ and neglect $\lambda$ corrections, so they are given
by the structure constant of the free $O(N)$ model, see
e.g.~\cite{Shachar:2022fqk}.

In the following we determine the fixed points of the beta
function~\eqref{eqn:beta4d} corresponding to conformal surface defects.

\subsection{Symmetry preserving defect}

The simplest fixed points of the beta function correspond to defects preserving
the full $O(N)$ symmetry, for which we should take $h_{ij} = h_N \delta_{ij}$.
Substituting this ansatz into the beta function and plugging the value of $\lambda =
\lambda_*$~\eqref{eqn:lambdas} we get two zeros
\begin{align}
  \frac{h_{N,+}}{2\pi} =
  \frac{6 \epsilon}{N+8} + O(\epsilon^2)\,,
  \qquad
  h_{N,-} =
  0\,.
  \label{eqn:hn4d}
\end{align}
The case $h_{N,-}$ is the trivial defect $D_0$, while $h_{N,+}$ is a new nontrivial
surface defect I call $D_N$. The RG flow between these two fixed points
is driven by the defect operator $\hat{\sigma} \sim \varphi^2_S$ (up to a
normalisation factor), which is relevant in $D_0$. At $D_N$ its
conformal dimension can be obtained from the derivative of the beta
function~\cite{Gubser:2008yx} and is irrelevant
\begin{align}
  \hat{\Delta}_{\hat{\sigma}}
  =
  2 + \beta'(h)|_{h_{N,+}}
  =
  2 + \frac{6 \epsilon}{N+8} + \dots
  \label{eqn:beta4dfpsym}
\end{align}

These defects are characterised by their sourcing of $\varphi^2_S$, which is
encoded in the dCFT coefficient $a_{\varphi^2}$ appearing
in~\eqref{eqn:coeffaphi2}. We read it from~\eqref{fig:renormalisationh4dl0a}
\begin{align}
  a_{\varphi^2}
  =
  -\frac{N h_{N,+}}{16 \pi^3 \sqrt{C_{\varphi^2_S}}}
  =
  - \frac{3 \sqrt{N} \epsilon}{\sqrt{2} (N+8)} + O(\epsilon^2)\,,
  \qquad
  C_{\varphi^2_S} = \frac{N}{8\pi^4} + O(\epsilon)\,.
  \label{eqn:aphi4d}
\end{align}

\subsection{Breaking to $O(p) \times O(N-p)$}
\label{sec:4dbreak}

A natural generalisation is to allow surface operators to break the $O(N)$
symmetry, and we start by discussing the breaking to the subgroup $O(p)
\times O(N-p)$. It turns out that, at this order in the $\epsilon$-expansion
there are no fixed points breaking more symmetries, so this is already the most
general case (we return to this question in section~\ref{sec:generalbreaking}).

To simplify our analysis, we use the $O(N)$ symmetry to
diagonalise $h_{ij}$ to its set of eigenvalues, and take $p$ eigenvalues of
value $h_p$ and $N-p$ eigenvalues of value $h_{N-p}$.
Decomposing the beta function~\eqref{eqn:beta4d} for the diagonal elements $h_p$ and $h_{N-p}$ of
$h_{ij}$, the vanishing of the beta functions reduces to%
\footnote{
One can also decompose the beta function in terms of irreducible
representations of $O(N)$, but it does not lead to a factorisation of the
equations because of the explicit symmetry breaking.
}
\begin{align}
  \beta_{h_p} &=
  - \epsilon h_p
  + \frac{\lambda}{(4\pi)^2} \frac{(p+2) h_p + (N-p) h_{N-p}}{3}
  + \frac{h_p^2}{2\pi}
  + \dots = 0\,,\\
  \beta_{h_{N-p}} &=
  - \epsilon h_{N-p}
  + \frac{\lambda}{(4\pi)^2} \frac{p h_{p} +  (N+2-p) h_{N-p}}{3}
  + \frac{h_{N-p}^2}{2\pi}
  + \dots = 0\,,
\end{align}
with the ellipsis containing terms subleading in $\epsilon$.

\paragraph{The fixed points.}
The solutions to these equations are easily obtained. The beta functions are
exchanged under the relabelling $p \to N-p$, so $\beta_{h_p} -
\beta_{h_{N-p}}$ is odd with respect to that symmetry and is given by
\begin{align}
  \beta_{h_p} - \beta_{h_{N-p}}
  =
  (h_p - h_{N-p})\left( \frac{h_p + h_{N-p}}{2\pi} - \epsilon +
  \frac{2 \lambda}{3 (4\pi)^2} + \dots \right)
  = 0\,.
\end{align}
In the symmetry breaking case, we require $h_p \neq h_{N-p}$.
This fixes $h_p + h_{N-p}$, and substituting into $\beta_{h_p} +
\beta_{h_{N-p}}$ we can solve for $h_p - h_{N-p}$. The solution is given by the
pair $\left\{ h_{p,+}, h_{N-p,-} \right\}$, where
\begin{align}
  \frac{h_{p,\pm}}{2\pi} &=
  \frac{(N + 3 - p) \pm \Delta}{N+8}
  \epsilon\,, \qquad
  \Delta^2 = 9 - p (N-p)\,.
  \label{eqn:4dsurfacebreakingsol}
\end{align}
The choice of sign for the square root gives rise to 2 different solutions for
any given $p$, and for simplicity here we used the freedom to relabel $p \to
N-p$ to choose a definite sign.  As a consistency check, we recover the symmetry
preserving defect $D_N$ defined in~\eqref{eqn:hn4d} by setting $p=N$, and the trivial
defect $D_0$ by setting $p = 0$.

The properties of these fixed points are controlled by the sign of the
discriminant $\Delta^2$. For $N$ below the critical value $N_c = 6$, the
discriminant is positive and the $h$'s are real numbers: these are fixed points
of the dRG flow preserving an $O(p) \times O(N-p)$ symmetry, and they correspond
to conformal defects which I call $D_p$.

Above the critical bound $N > N_c$, there are values of $p$ for which the
discriminant is negative. This happens for $1 < p < N-1$ for $6 < N \le 10$, and
$0 < p < N$ for $N > 10$ (for integer $p$). In that case, \eqref{eqn:4dsurfacebreakingsol}
describes complex fixed points of the beta function. In a unitary theory, the
$h$'s remain real along dRG flows, so these fixed points are not reached by a
dRG flow (they may however lead to walking behavior as in~\cite{Gorbenko:2018ncu}).

At certain values of $p(N)$ the discriminant vanishes exactly, so that $h_{p,+} =
h_{p,-}$ and the defects $D_p$ and $D_{N-p}$ become degenerate. This happens at
integer values of $p$ for $N=6$ ($p=3$) and $N = 10$ ($p=1,9$). In fact, for
$N=6$ one can check that the fixed points~\eqref{eqn:4dsurfacebreakingsol} have
the same values of $h$ for $p = 3,4,5,6$, so that all four defects $D_p$ become
degenerate and coincide with the symmetry preserving defect~\eqref{eqn:hn4d}.
As far as I know this is the first example of a simultaneous collision of four
fixed points! It would be interesting to see if this degeneracy is lifted at
subleading orders in $\epsilon$.

We can gather more information about the structure of the dRG flow from the beta
function~\eqref{eqn:beta4d}. The eigenvalues of its jacobian give the conformal
dimension of defect operators involved in the flow. There are four different
eigenvalues, which correspond to four defect operators:
\begin{table}[h]
  \centering
  \begin{tabular}{cccc}
    \toprule \midrule
    Operator & $O(p)$ & $O(N-p)$ & $\hat{\Delta} - 2$\\ \midrule
    $\hat{\cO}_{S,+}$ & $S$ & $S$ &
    $\frac{\epsilon}{2(N+8)}\left[ N + \sqrt{N^2 + 8(N-2p) \Delta + 16 \Delta^2}
  \right]$ \\
  $\hat{\cO}_{S,-}$ & $S$ & $S$ &
    $\frac{\epsilon}{2(N+8)}\left[ N - \sqrt{N^2 + 8(N-2p) \Delta + 16 \Delta^2}
  \right]$ \\
  $\hat{\cO}_{T,+}$ & $T$ & $S$ & $\frac{N-2p+2\Delta}{N+8} \epsilon$\\
  $\hat{\cO}_{T,-}$ & $S$ & $T$ & $-\frac{N-2p+2\Delta}{N+8} \epsilon$\\
  \bottomrule
  \end{tabular}
  \caption{List of defect operators involved in the dRG flow for the defect
    $D_p$, their representation (singlet $S$/traceless symmetric $T$) and their
    conformal dimension $\hat{\Delta}$.  Note that $\hat{\cO}_{S,+}$ is absent
    when $p = 0$, $\hat{\cO}_{S,-}$ when $p = N$, $\hat{\cO}_{T,+}$ when $p < 2$
    and $\hat{\cO}_{T,-}$ is absent when $N-p < 2$.}
  \label{tab:dopsDp}
\end{table}

\begin{figure}[tb]
  \centering
  \includegraphics[scale=1]{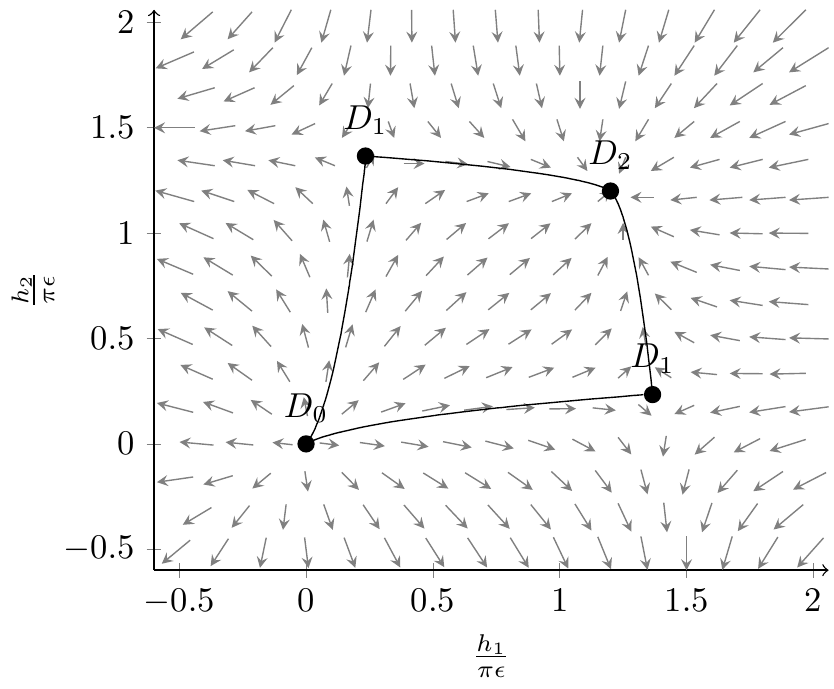}
  \caption{Example of a defect RG flow for surface defects in the $O(2)$ model.
    The vector field is $-\beta(h_1,h_2)$ given in terms of the eigenvalues
    $h_1, h_2$. There is a $\mathbb{Z}_2$ symmetry exchanging $h_1
    \leftrightarrow h_2$. The 3 fixed points are $D_0$, $D_1$ and $D_2$, their
    values of $h$'s are given in~\eqref{eqn:4dsurfacebreakingsol}. The black
  lines indicate the stable manifold.}
  \label{fig:rgflowex}
\end{figure}

The operators relevant for assessing the stability of the fixed points are $\hat{\cO}_{S,\pm}$.
Notice that for $N \le 6$, $\hat{\cO}_{S,-}$ is a relevant defect operator (i.e.
$\hat{\Delta} < 2$) present in all defects $D_p$ except the symmetry
preserving defect $D_N$. Therefore we expect a dRG flow between the defects
$D_0 \to D_p \to D_{N-p} \to D_N$, with $D_N$ the IR fixed point.
As an example, see figure~\ref{fig:rgflowex} illustrating the dRG flow for $N=2$.

For $N \ge 6$ things are more intriguing. The operator $\hat{\cO}_{S,-}$ becomes
marginal when $\Delta = 0$. This suggests that $\hat{\cO}_{S,-}$ is responsible
for the flow between $D_p \to D_{N-p}$ ($p < N-p$), and it becomes marginal
because the fixed points collide. Note that between $6 \le N \le 10$, we have
symmetry breaking unitary defects for $p = 1$ and $p = N-1$, and for the defect
$D_{N-1}$ the operator $\hat{\cO}_{S,-}$ is \textit{irrelevant}, so it
corresponds to an IR fixed point of the dRG flow.

Finally, there are the defect operators $\hat{\cO}_{T,\pm}$, which trigger
dRG flows with an explicit symmetry breaking (e.g. the flow $D_0 \to D_1$ in
figure~\ref{fig:rgflowex}). It's interesting to note that the symmetry
preserving defect $D_N$ contains the operator $\hat{\cO}_{T,+}$ which becomes
relevant for $N > 6$
\begin{align}
  \hat{\Delta}_{T,+} - 2
  = 
  -\frac{N - 6}{N+8} \epsilon + \dots
\end{align}
This is easy to understand at large $N$. The dRG
flow mixes the singlet and symmetric traceless representations of $h_{ij}$
through the quadratic term $h_{ik} h_{kj} \sim (h_S^2 + \frac{1}{N} h_T^2)
\delta_{ij} + \dots$ of the beta function~\eqref{eqn:beta4d}. In the large $N$
limit this interaction term vanishes and the representations decouple, leaving
$\hat{\cO}_{T,+}$ the symmetric traceless part of $[\varphi^i \varphi^j]$, which
is a relevant operator.

Note that even though $\hat{\cO}_{T,+}$ is a relevant operator in this case, it
involves an explicit symmetry breaking so does not signal an instability of the
fixed points.

\paragraph{The conformal manifold.}
An interesting feature of these symmetry breaking defects is the presence of a
conformal manifold. 
Any choice of matrix $h_{ij}$ with eigenvalues $h_p$ and $h_{N-p}$ as above
defines an equivalent defect $D_p$ (i.e. they are related by an $O(N)$ transformation).
The space of such matrices is the Grassmanian
\begin{align}
  \mathbf{Gr}_p(N)
  =
  \frac{O(N)}{O(p) \times O(N-p)}\,,
\end{align}
which is interpreted as the conformal manifold for the dCFT.  The breaking of
symmetry implies the existence of local operators $\bO^{a m}$ ($a = 1,\dots,p$
and $m = p+1,\dots,N$) in the dCFT which arise as contact terms for the broken
$O(N)$ currents
\begin{align}
  \partial_\mu j^{\mu a m}(x) D_p
  =
  D_{p}[\bO^{a m}(x_\parallel)] \delta^{(d-2)}(x_\perp)\,,
  \qquad
  \bO^{a m}
  =
  (h_{p,+}+h_{N-p,-}) M^{\epsilon} \varphi^a \varphi^{m}\,.
\end{align}
These operators---sometimes called tilt operators---are exactly marginal and
generate translations on the conformal manifold. Because of their geometric
interpretation, their correlators are constrained by integrated Ward
identities~\cite{Drukker:2022pxk} (see also~\cite{Herzog:2023dop}).

We note that the combination of eigenvalues appearing in the tilt operator has
the simple expression
\begin{align}
  h_{p,+} + h_{N-p,-}
  =
  \frac{2 (N+6)}{(N+8)} \pi \epsilon + \dots
\end{align}

\subsection{General breaking of symmetries}
\label{sec:generalbreaking}

One can also consider more general dRG flows, where we allow for the $O(N)$
symmetry to be broken to a subgroup $\prod_l O(p_l)$. It is a simple exercise to
write and solve the beta functions for $l=3$, which reveals that
there is no new fixed point with 3 different eigenvalues $h_l$ at this order in
$\epsilon$.

This follows from a simple counting argument. 3 quadratic equations can have at
most 8 solutions, and all of them are already accounted for by degenerate cases
where two eigenvalues coincide: In addition to the trivial solution $h = 0$ and
the symmetry preserving defect $h_{p_1} = h_{p_2} = h_{p_3}$, there are 3
conditions for degeneracy ($h_1 = h_2$ and permutations) and in each case we
have 2 nontrivial solutions. Therefore there cannot be new fixed points with
$l = 3$ symmetry breaking at this order in perturbation theory, and by induction
$l > 3$ can be excluded as well.

It would be interesting to see if subleading corrections to $h$ lift the
degeneracies and lead to defects with more complicated breakings of symmetry.

\section{Large $N$ expansion}
\label{sec:largen}

To understand the fate of surface operators across dimensions and especially at
$d = 3$, a convenient tool is the large $N$ expansion. From the analysis near
$d=4$, we expect the symmetry preserving defects $D_N$ to be the most relevant in the IR,
so in this section we focus on these and show that they can be reliably studied
away from $d = 4$.

The standard way to approach the large $N$ limit of the $O(N)$ model is to use
the Hubbard-Stratonovich transformation, see~\cite{Moshe:2003xn} for a review.
Following~\cite{Fei:2014yja}, we rewrite the action~\eqref{eqn:S4d} with an
auxiliary field $\sigma$ as
\begin{align}
  S
  &=
  \int \diff^d x \left[
    \frac{1}{2} \left( \partial_\mu \varphi^i \right)^2
    + \frac{1}{2 \sqrt{N}} \sigma \varphi^i \varphi^i
    - \frac{3 M^{d-4}}{2 N \lambda} \sigma^2
  \right]\,.
\end{align}
Upon integrating out $\sigma$ one recovers the original action~\eqref{eqn:S4d}.
At large $N$, the interaction term $\sigma \varphi^2$ promotes
$\sigma$ to a dynamical field by generating a effective propagator from
an infinite sum of bubble diagrams (the solid black line is the propagator for
$\sigma$ and $\bullet$ is the vertex $\sigma \varphi^2$)
\begin{align}
  \FDprops = \FDpropphia + \FDpropphib + \dots
\end{align}
These diagrams assemble into a geometric series, and resumming one finds the
effective propagator
\begin{align}
  \vev{\sigma(x_1) \sigma(x_2)}
  =
  \frac{C_{\sigma}}{|x_{12}|^{2\Delta_\sigma}}\,,
  \qquad
  C_{\sigma}
  =
  \frac{2^{d+2} \Gamma\left( \frac{d-1}{2} \right) \sin\left( \frac{\pi d}{2}
  \right)}{\pi^{3/2} \Gamma\left( \frac{d}{2} -2 \right)}
  + O(N^{-1})\,.
  \label{eqn:2ptsfuncs}
\end{align}
The conformal dimension is $\Delta_\sigma = 2 + \gamma_\sigma$ with the
anomalous dimension subleading in $N$ at large $N$ and given by
\begin{align}
  \gamma_{\sigma} =
  \frac{1}{N}
  \frac{4 \Gamma\left( d \right) \sin\left( \frac{\pi d}{2} \right)}{\pi
    \Gamma\left( \frac{d}{2}+1 \right)\Gamma\left( \frac{d}{2}-1 \right)}
    + O(N^{-2})\,.
  \label{eqn:gammas}
\end{align}
In the following we also need the 3-point function of $\sigma$ obtained
in~\cite{Petkou:1994ad,Petkou:1995vu,Goykhman:2019kcj}
\begin{align}
  \vev{\sigma(x_1)\sigma(x_2)\sigma(x_3)}
  &=
  \frac{C_{\sigma \sigma \sigma}}{|x_{12}|^{\Delta_\sigma}|x_{13}|^{\Delta_\sigma}
  |x_{23}|^{\Delta_\sigma}}\,,\\
  C_{\sigma \sigma \sigma}
  &=
  -\frac{1}{\sqrt{N}}
  \frac{\Gamma\left( 3-\frac{d}{2} \right)}{\Gamma\left( d-3 \right)}
  \left(\frac{2^{d-1} \Gamma\left( \frac{d-1}{2} \right) \sin \left(\frac{\pi
  d}{2} \right)}{\pi^{3/2}} \right)^3 + O(N^{-3/2})\,.
  \label{eqn:largenCsss}
\end{align}

In this description, symmetry preserving defects are given by the integral of
$\sigma$
\begin{align}
  D = \exp\left( -\int \diff^2 \tau h\, \sigma(\tau) \right)\,.
  \label{eqn:defectlargen}
\end{align}
The beta function for the coupling $h$ is given
by~\eqref{eqn:defectbetafunction}. Solving for its zero we find the fixed point
corresponding to the defect $D_N$ for any $d$
\begin{align}
  \beta_h
  =
  \gamma_\sigma h
  + \pi {C_{\sigma \sigma}}^{\sigma} h^2 + \dots\,,
  \quad \Rightarrow \quad
  h_* = - \frac{\gamma_\sigma}{{C_{\sigma \sigma}}^\sigma \pi} + \dots\,.
  \label{eqn:betalargen}
\end{align}

The structure of the dRG flow is easy to understand and follows from the sign of
the anomalous dimension $\gamma_\sigma$. When $2 < d < 4$, the anomalous
dimension is negative $\gamma_\sigma < 0$, so the operator $\sigma$ is
marginally relevant and leads to a dRG flow from the trivial defect to
$D_N$. The $O(N)$ model is still defined for $4 < d < 6$~\cite{Fei:2014yja}
but in that case $\gamma_\sigma > 0$, so $\sigma$ is a marginally
irrelevant perturbation. The defect $D_N$ may then be thought of as a UV fixed
point of the dRG flow ending at the trivial defect.

\begin{figure}[h]
  \centering
  \begin{subfigure}[b]{0.5\textwidth}
    \begin{tikzpicture}[baseline=(current bounding box.center)]
  \begin{axis}[
    domain=2:6,
    axis y line=left,
    axis x line=middle,
    axis line style={->},
    xlabel={$d$},
    ylabel={$\sqrt{N} a_{\varphi^2}$}
  ]
    \addplot[MidnightBlue,thick] table{data/aphi.dat};
    \addplot[MidnightBlue,thick] table{data/aphi2.dat};
  \end{axis}
\end{tikzpicture}
  \end{subfigure}
  \hspace{0.1\textwidth}
  \begin{subfigure}[b]{0.3\textwidth}
  \renewcommand{\arraystretch}{1.5}
  \begin{tabular}{cc}
    \toprule \midrule
    $d$ & $\sqrt{N} a_{\varphi^2}$\\
    \midrule
    $2+\epsilon$ & $\epsilon$\\
    $3+\epsilon$ & $-\frac{8}{3 \pi \epsilon}$\\
    $4-\epsilon$ & $-\frac{3|\epsilon|}{\sqrt{2}}$\\
    $5$ & $-\frac{16 \sqrt{2}}{5\pi}$\\
    $6-\epsilon$ & $-\frac{10}{3} \sqrt{\frac{2 \epsilon}{3}}$\\
      \bottomrule
  \end{tabular}
  \renewcommand{\arraystretch}{1}
  \end{subfigure}
  \caption{On the left, a plot of $a_{\varphi^2}(d)$ to leading order at large
$N$. On the right, the corresponding values near integer dimensions.}
  \label{fig:aphi}
\end{figure}
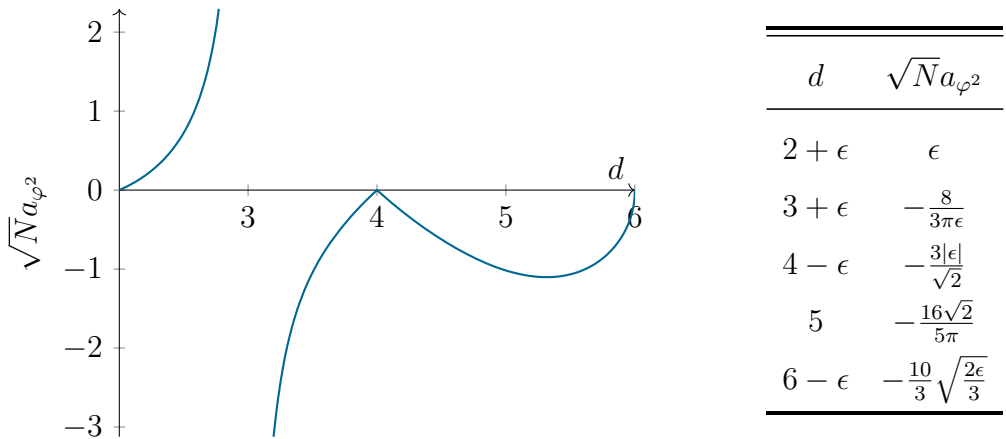

A short calculation shows that the 1-point function of $\sigma$ is given by
\begin{align}
  \vev{\sigma(x) D}
  =
  - h_* \int \diff^2 \tau \vev{\sigma(x) \sigma(\tau)}
  + \dots
  =
  - \frac{\pi C_\sigma h_*}{x^2}\,.
\end{align}
The normalisation independent quantity associated to $h_*$ is $a_{\varphi^2}$,
which we obtain by dividing by the normalisation of
$\sigma$~\eqref{eqn:2ptsfuncs} and comparing to~\eqref{eqn:coeffaphi2}
\begin{align}
  a_{\varphi^2}
  =
  - \pi \sqrt{C_\sigma} h_*
  =
  -\frac{1}{\sqrt{N}}
  \frac{2^{1+\frac{d}{2}}(d-1)}{\pi^{\frac{3}{4}} d(d-3)}
  \sqrt{\frac{\Gamma\left( \frac{d-1}{2} \right) \sin\left(
  \frac{\pi d}{2} \right)}{\Gamma\left( \frac{d-4}{2} \right)}}
  + O(N^{-3/2})\,.
  \label{eqn:aphi2largenres}
\end{align}
A plot of $a_{\varphi^2}$ is presented in figure~\ref{fig:aphi}, and its values at
integer dimensions are given in table~\ref{fig:aphi}.
At dimensions $d = 2,4,6$, it has zeros because the anomalous dimension
$\gamma_\sigma$~\eqref{eqn:gammas} vanishes at large $N$ for even integer
dimensions.
In particular near $d=4$, it behaves as $-|\epsilon|$ since both $\gamma_\sigma$
and $C_{\sigma\sigma\sigma}$ change sign, and the coefficient matches with the
large $N$ limit of the result from the $\epsilon$-expansion~\eqref{eqn:aphi4d}
as expected.

An interesting feature is the divergence at $d=3$, which follows from
the vanishing of the structure constant $C_{\sigma\sigma\sigma}$ at large
$N$~\cite{Petkou:1994ad,Petkou:1995vu,Goykhman:2019kcj}. This signals a
breakdown of the small $h_*$ approximation leading to further corrections of the beta
function~\eqref{eqn:betalargen}. We discuss two cases.

The first possibility is that the zero of the beta function becomes of order
$O(N^0)$. The leading contribution to the renormalisation of the coupling $h$
comes from the diagram
\begin{align}
  \FDfi
  \sim
  \int_{\bR^2} \diff^2 \tau_{1,2,3}
  \vev{\sigma(x) \sigma(\tau_1) \sigma(\tau_2) \sigma(\tau_3)}_{connected}
  \sim \frac{h^3}{N} \frac{A}{x^2} + \dots
\end{align}
for some $A$. If $A$ is nonzero and the diagram contributes to the beta
function, then we expect $\beta \sim \gamma_\sigma h + \frac{B h^3}{N} +
\dots = 0$ for some $B$, and since $\gamma_\sigma \sim N^{-1}$ we would find two
nontrivial fixed points $h_* \sim O(N^0)$.

Doing this calculation explicitly seems difficult because the 4-point function
of $\sigma$ is only known partially at large $N$~\cite{LANG1993573}. However one
might suspect that the fixed points are absent ($B = 0$) for two reasons. First,
there have been explicit studies of boundary/interface fixed points at low values of
$N$ (e.g.~\cite{PhysRevE.72.016128,Liendo:2012hy,Gliozzi:2015qsa}) with no
indications of such new fixed points.
Second, a simpler version of this question is to restrict to $d = 4-\epsilon$,
where the Feynman diagrams can be evaluated more easily. This is presented in
appendix~\ref{sec:allloop} and we verify that $B|_{d = 4-\epsilon} = 0$, which
suggests that perhaps also $B|_{d=3} = 0$.

The second possibility is that the zero of the beta function becomes of order
$\sqrt{N}$. In this regime the behavior of the fixed points is well-understood
from the large $N$ analysis
of~\cite{AJBray_1977,ohno,McAvity:1995zd,Giombi:2020rmc,Krishnan:2023cff} for an
interface in $d=3$. There are two symmetry preserving fixed points, known in the
literature as the special and ordinary transitions. The special transition
corresponds to the trivial interface and is characterised by
$a_{\varphi^2,sp} = 0$. The ordinary transition on the other hand is a
nontrivial interface and is characterised by~\cite{McAvity:1995zd}
\begin{align}
  a_{\varphi^2,ord}
  =
  -\frac{\pi \sqrt{N}}{8} + O(N^{-1/2})\,.
\end{align}
Therefore I expect that the divergence of $a_{\varphi^2}$ at $d = 3$ should be
interpreted as a change in scaling with respect to $N$, and that $D_N$ becomes
the ordinary fixed point.

Finally, note that there is a third fixed point known as the extraordinary
(or normal) transition. That interface is characterised by a nonzero one-point
function for $\varphi^i$ breaking the $O(N)$ symmetry down to $O(N-1)$. Below
$d = 3$, $a_{\varphi^2}$ becomes positive (or equivalently
$h_* < 0$), which can be interpreted as sourcing a negative mass for
$\varphi^2$.
This signals an instability at $\phi^i = 0$, and including a defect coupling
$u^i \phi^i$ in~\eqref{eqn:defectlargen}, we expect the stable IR fixed point to
have nonzero $u^i$, leading to the extraordinary fixed point.

\section{\texorpdfstring{$\epsilon$}{Epsilon}-expansion in $d=6-\epsilon$}
\label{sec:pert6d}

The critical $O(N)$ model can be analytically continued to the range $4 < d <
6$, where it is described by the (perturbative) UV fixed point of the
action~\eqref{eqn:S4d}. It can be studied using an $\epsilon$-expansion at
$d=6-\epsilon$ from the action~\cite{Fei:2014yja}
\begin{align}
  S = \int \diff^{6-\epsilon} x
  &\left[ \frac{1}{2} (\partial_\mu \varphi^i)^2
    + \frac{1}{2} \left( \partial_\mu \sigma \right)^2
    + \frac{M^{\frac{\epsilon}{2}} g_1}{2} \sigma \varphi^i \varphi^i
    + \frac{M^{\frac{\epsilon}{2}} g_2}{6} \sigma^3
  \right]\,.
  \label{eqn:S6d}
\end{align}
This theory is conformal when setting the coupling constants to their critical
values $g_1^*, g_2^*$, which have been calculated in~\cite{Fei:2014yja}%
\footnote{
  These are asymptotic series at large $N$. The behavior of the fixed point at
  finite $N$ is discussed in~\cite{Fei:2014yja}.
}
(see also~\cite{Fei:2014xta,Gracey:2015tta,Kompaniets:2021hwg,Borinsky:2021jdb}
for subleading corrections up to $O(\epsilon^5)$)
\begin{equation}
  \begin{aligned}
    g_1^* &= \sqrt{\frac{6 (4\pi)^3 \epsilon}{N}}
    \left( 1 + \frac{22}{N} + \frac{726}{N^2} - \frac{326180}{N^3} + \dots
      \right)\,,\\
    g_2^* &= \sqrt{\frac{6 (4\pi)^3 \epsilon}{N}}
    \left( 6 + \frac{972}{N} + \frac{412596}{N^2} + \frac{247346520}{N^3} + \dots
    \right)\,.
  \end{aligned}
  \label{eqn:g6dsol}
\end{equation}

In this section we study surface operators in the $d = 6-\epsilon$ description.
Doing so provides another nontrivial check of the large $N$ results obtained in
section~\ref{sec:largen}. Perhaps more importantly, because the dimension of a
free scalar field is $\Delta_\varphi = 2$ in $d = 6$, we can reliably study dRG
flows of surface operators with a coupling to $\varphi^i$ using perturbation
theory
\begin{align}
  D
  =
  \exp\left[ -\int \diff^2 \tau \left( h \sigma + u^i \varphi^i \right) \right]\,.
  \label{eqn:surface6d}
\end{align}
When $u \neq 0$ these surface operators preserve an $O(N-1)$ symmetry and are
relevant for describing the extraordinary transition in $d = 3$.

Unlike our treatment in section~\ref{sec:pert4d}, here to any order in $g_1$ and
$g_2$ the number of Feynman diagrams contributing to the renormalisation of the
defect is finite, and we can treat $h, u$ nonperturbatively (this is similar
to~\cite{Cuomo:2021kfm}). The beta functions
for the defect couplings involve a term linear in $\epsilon$ from the classical
dimension of the fields, so for consistency one should calculate the beta
functions to order $g^2 \sim \epsilon$.  The diagrams contributing to that order
can be found in tables~\ref{tab:feyn6dh} and~\ref{tab:feyn6dn}.  A
straightforward calculation detailed in appendix~\ref{app:pert6d} leads to the
beta functions
\begin{equation}
\begin{aligned}
  \beta_h &=
  \left(\Delta_\sigma - 2 \right) h
  - \frac{g_1 u^2 + g_2 h^2}{(4\pi)^2}
  - \frac{2 g_1^2 h u^2 + g_1 g_2 h u^2 + g_2^2 h^3}{(4\pi)^4} + \dots\\
  \beta_{u^i} &=
  \left( \Delta_\varphi - 2 \right) u^i
  - \frac{2 g_1 h u^i}{(4\pi)^2}
  - \frac{g_1 (2 g_1 h^2 + g_1 u^2 + g_2 h^2) u^i}{(4\pi)^4} + \dots
\end{aligned}
  \label{eqn:beta6d}
\end{equation}
where the conformal dimensions are~\cite{Fei:2014yja}
\begin{align}
  \Delta_\sigma &= 
  2 - \frac{\epsilon}{2} + \frac{1}{(4\pi)^3} \frac{N g_1^2 + g_2^2}{12} + O(g^4)
  = 2 + \left( \frac{40}{N} + \frac{6800}{N^2} + \dots \right)
  \epsilon + O(\epsilon^2)\,,\\
  \Delta_\varphi &=
  2 - \frac{\epsilon}{2} + \frac{g_1^2}{6 (4\pi)^3} + O(g^4)
  =
  2 + \left( -\frac{1}{2} + \frac{1}{N} + \frac{44}{N^2} + \frac{1936}{N^3} + \dots \right)
  \epsilon + O(\epsilon^2)\,.
  \label{eqn:gammasigma}
\end{align}
The terms linear and quadratic in $h, u$ match the general
form~\eqref{eqn:defectbetafunction}, and in addition we have terms cubic 
in $h, u$ and quadratic in $g$.

The simplest fixed point is the trivial defect with $h = u = 0$.
Below we analyse the fixed points describing the symmetry preserving defect
$D_N$ as well as symmetry breaking defects.

Note that similar dRG flows were considered
in~\cite{Rodriguez-Gomez:2022gbz,Bolla:2023zny,SoderbergRousu:2023zyj}. Their
analysis involves a double-scaling limit where the bulk theory is not critical
such that diagrams contributing to the anomalous dimension of $\varphi, \sigma$
are suppressed.

\subsection{Symmetry preserving defect}

Setting $u=0$, the beta function for $h$ factorises. There are 3 solutions,
$h=0$ and two nontrivial fixed points at
\begin{align}
  h_{*} = -\frac{8 \pi^2}{g_2} \left( 1 \pm \sqrt{1 + 4 (\Delta_\sigma - 2)} \right)\,.
  \label{eqn:hstar6d}
\end{align}

The simpler fixed point to analyse is the negative root, which is a perturbative
fixed point analogous to those of sections~\ref{sec:pert4d}
and~\ref{sec:largen}. We find
\begin{align}
  h_*
  =
  \frac{16 \pi^2}{g_2} \left( \Delta_\sigma - 2 \right)
  =
  \frac{20}{3} \sqrt{\frac{2 \pi \epsilon}{3 N}} \left( 1 +
  \frac{8}{N} - \frac{4118}{N^2} + \dots \right)\,.
  \label{eqn:h6dsym}
\end{align}
This describes the symmetry preserving defect $D_N$, as can be shown by
comparing the value of $a_\varphi^2$ to the large $N$ result.
A short calculation of the 1-point function of $\sigma$ (see
table~\ref{tab:feyn6dh}) allows us to extract
\begin{align}
  a_{\varphi^2}
  =
  -\frac{10}{3} \sqrt{\frac{2 \epsilon}{3N}}
  \left(  1 + \frac{8}{N} - \frac{4118}{N^2} + \dots \right)\,,
\end{align}
which agrees with the leading large $N$ result~\eqref{eqn:aphi2largenres}, see
figure~\ref{fig:aphi}.

Since $\sigma$ is an irrelevant defect operator ($\Delta_\sigma > 2$) at the
trivial defect, one should interpret~\eqref{eqn:h6dsym} as a UV fixed point
for a dRG flow.  Indeed, we can check that $\hat{\sigma}$ is marginally
relevant at the fixed point
\begin{align}
  \hat{\Delta}_{\hat{\sigma}}
  =
  2 + \beta'(h)|_{h_*}
  =
  2 - \left(\frac{40}{N} + \frac{6800}{N^2} + \dots \right) \epsilon +
  O(\epsilon^2)\,.
\end{align}

The fixed point at the positive root of~\eqref{eqn:hstar6d} is more subtle. 
We have $h_* \sim g_2^{-1}$ for a large negative value of
$h_*$. Since the combination $h_* g_* \sim O(\epsilon^{0})$ is not a small
parameter, the beta function may receive additional contributions and the
validity of the solution cannot be established in our approximation. 
One may expect that it disappears when including subleading corrections to the
beta function~\eqref{eqn:beta6d}.  For instance, when using the Padé approximant
\begin{align}
  \beta_h \supset
  -\frac{g_2 h^2}{(4\pi)^2}
  - \frac{g_2^2 h^3}{(4\pi)^4}
  + \dots
  \sim
  -\frac{h}{1 - \frac{g_2 h}{(4\pi)^2}}\,,
\end{align}
the beta function only has the fixed points $h = 0$ and~\eqref{eqn:h6dsym}.

\subsection{Symmetry breaking defect}

We now turn to solutions with $u \neq 0$, which breaks the $O(N)$ symmetry to
$O(N-1)$. It's convenient to write $u^i = u n^i$, with $n^i$ a unit length
vector parametrising the symmetry breaking.
For nonzero $u$ there are 3 solutions to the beta functions, and only one
can be studied reliably using perturbation theory (the other two have $h_*, u_*
\sim \sqrt{N/\epsilon}$ with $h_* < 0$). Taking $h_*, u_* \sim \sqrt{\epsilon}$ as an ansatz, it
is easy to solve the beta functions to find
\begin{align}
  h_* &=
  \frac{\Delta_\varphi - 2}{2 \tilde{g}_1} +
  O(\epsilon^{3/2})\,,\\
  u_*^2 &=
  \frac{\pi^2 (\tilde{g}_1^2-\frac{3\epsilon}{4\pi})(N \tilde{g}_1^3+ \tilde{g}_1
  \tilde{g}_2^2 - \tilde{g}_1^2 \tilde{g_2} + (\tilde{g}_2 - 2 \tilde{g}_1)
  \frac{3 \epsilon}{4\pi})}{9\tilde{g}_1^3}
  + O(\epsilon^{3/2})\,.
\end{align}
Plugging in the value of $g$~\eqref{eqn:g6dsol} we get
\begin{align}
  h_* &= -\frac{1}{2} \sqrt{\frac{\pi \epsilon}{6N}} \left[
  1 - \frac{24}{N} - \frac{286}{N^2} + \frac{346024}{N^3} + \dots \right]\,,\\
  u_*^2 &=
  - \frac{\pi N \epsilon}{12} \left[ 3 + \frac{356}{N} + \frac{184 652}{N^2} +
  \frac{117 474 208}{N^3} + \dots \right]\,.
  \label{eqn:hu6dsymbreak}
\end{align}
$u_*^2$ is negative since $\tilde{g}_1^2 < \frac{3 \epsilon}{4\pi}$. This is
clear from the asymptotic expression at large $N$ and can be checked for finite
$N \gtrsim 1038.27$ using the expressions for $g$ given in~\cite{Fei:2014xta}.
The solution is then a complex fixed point which corresponds to a nonunitary dCFT.
As already mentionned in section~\ref{sec:4dbreak}, such fixed points are not
reached by unitary dRG flows.

Note that $u_* \sim \sqrt{N}$ has the correct scaling to describe the normal
transition in $d = 3$.  I report here the eigenvalues of the jacobian at the
fixed points
\begin{align}
  \epsilon \left[ \frac{3 \pm \sqrt{3}}{2} + \frac{227 \pm
  \frac{437}{\sqrt{3}}}{N} + \frac{100 288 \pm \frac{557 024}{3\sqrt{2}}}{N^2} +
\dots\right]\,.
\end{align}
Both of these are real and positive. In addition to these operators, the dCFT
contains a tilt operator $\bO^i = u_* \varphi^i$ associated to the breaking of
symmetry $O(N) \to O(N-1)$. Since $u_*$ is imaginary, these operators have
negative norm.

\section{The anomaly coefficients}
\label{sec:anomaly}

Having obtained the fixed points describing conformal surface operators in the
previous sections, I now turn to the calculation of their anomaly
coefficients~\eqref{eqn:anomalycoeff}. Anomaly coefficients have been studied in
the past in many contexts, and it turns out that most of the results we need are
already known. The coefficient $a$ for a surface operator of the
form~\eqref{eqn:defectCPT} was obtained in~\cite{Jensen:2015swa,Shachar:2022fqk}
using conformal perturbation theory, and their calculation can be applied to
our defects as well, with the exception of $d = 3$ where conformal perturbation
theory no longer applies (in $d = 3$ however the coefficient $a$ was
obtained independently in~\cite{Giombi:2020rmc,Krishnan:2023cff}). The rest of
the coefficients were calculated previously for surfaces in
6d~\cite{Henningson:1999xi,Gustavsson:2004gj,Drukker:2020dcz}, and as I discuss
below their calculation can be extended to surfaces in arbitrary dimensions as
well.

\paragraph{Surface in $d = 6 - \epsilon$.}
The simplest case to consider is surfaces in 6d, which couple directly to the
scalar fields $\sigma$ and $\varphi^i$ as in~\eqref{eqn:surface6d}. Their conformal
anomaly can be obtained by calculating the expectation value of a surface
operator over the arbitrary surface $\Sigma \subset \bR^{6-\epsilon}$ and
placing the theory on a curved background of metric $G$. In perturbation theory,
the leading contribution to their conformal anomaly is given by an integrated
propagator of the form
\begin{align}
  \frac{1}{2} \int_\Sigma \vol_\Sigma(\tau_1) \vol_\Sigma(\tau_2)
  u^i(\tau_1) u^j(\tau_2) \vev{\varphi^i(\tau_1) \varphi^j(\tau_2)}_G
  + O(u^3)\,,
  \label{eqn:intpropanomaly}
\end{align}
and we're interested in the $\log\tilde\epsilon$ divergences of that integral,
where $\tilde{\epsilon}$ is a UV cutoff.
These divergences come from coincident points $\tau_1 \to
\tau_2$ and obey the structure of a conformal anomaly~\eqref{eqn:anomalycoeff}.
They can be calculated from the small distance expansion of the curved space
propagator for a conformal scalar in 6d, which reads~\cite{Gustavsson:2004gj}
(in normal coordinates about the origin)
\begin{align}
  \vev{\varphi^i(0) \varphi^j(\xi)}_G
  =
  \frac{C_{d,1} \delta^{ij}}{|\xi|^4}
  \left[ 1 + \frac{1}{3} P_{\mu\nu} \xi^\mu \xi^\nu + \dots \right]\,,
  \label{eqn:conformalprop}
\end{align}
with $C_{d,1}$ the canonical normalisation for a scalar field in $d$ dimensions
defined in~\eqref{eqn:fourier} and $P_{\mu\nu} = \frac{1}{d-2} \left( R_{\mu\nu} -
\frac{2}{d-1} R G_{\mu\nu} \right)$ is the Schouten tensor.

The corresponding anomaly coefficients were obtained
in~\cite{Henningson:1999xi,Gustavsson:2004gj,Drukker:2020dcz} and are given by
\begin{align}
  a = 0\,, \qquad
  b_1 = -\frac{\pi^2 C_{d,1} u^2}{2}\,, \qquad
  b_2 = \frac{\pi^2 C_{d,1} u^2}{3}\,, \qquad
  c = \pi^2 C_{d,1} u^2\,.
  \label{eqn:anomalycoeffres}
\end{align}

The vanishing of $a$ to leading order has a simple explanation: the coefficient
$a$ measures the conformal anomaly associated to a change in topology from the
plane to the sphere, which is zero because the 2-point function of a (conformal)
scalar appearing in the integrand of~\eqref{eqn:intpropanomaly} is conformally
invariant.%
\footnote{
  This is no longer true if the operators have anomalous dimensions, in which
  case the coefficient receives subleading contributions.
}

For a surface operator obtained by integrating $u_i
\cO^i$ as in~\eqref{eqn:defectCPT}, \cite{Jensen:2015swa,Shachar:2022fqk} shows that the
leading contribution to $a$ instead comes at order $O(\epsilon u^2)$ and is
given by
\begin{align}
  a = \frac{\pi^2}{6} (\Delta_\cO - 2) C_{\cO} u^2 + O(u^4)\,,
  \label{eqn:anomalya}
\end{align}
where $C_\cO$ is the normalisation constant appearing in the 2-point function of
$\cO$.

From these results we can read the anomaly coefficients for surfaces
defects~\eqref{eqn:surface6d} at $d = 6 - \epsilon$. Adding the contribution of
each scalar, we get to leading order,
\begin{align}
  a = \frac{(\Delta_\sigma - 2)h_*^2 + (\Delta_\varphi - 2)u_*^2}{24\pi}\,,
  \qquad
  b_1 = -\frac{h_*^2 + u_*^2}{8\pi}\,,
  \qquad
  b_2 = \frac{h_*^2 + u_*^2}{12 \pi}\,,
  \qquad
  c = \frac{u_*^2}{4\pi}\,,
\end{align}
where the conformal dimensions are given in~\eqref{eqn:gammasigma} and the
values of the fixed points are respectively in~\eqref{eqn:h6dsym}
and~\eqref{eqn:hu6dsymbreak} for the symmetry preserving and breaking defects.
We note that the contribution from the symmetry breaking term
in~\eqref{eqn:anomalycoeff} is here $(\partial n)^2 = \partial_a n^i
\partial_b n^i h^{ab}$ with $h_{ab}$ the induced metric on $\Sigma$ and $n^i$ is
the unit vector parallel to $u^i$.

Focussing on the symmetry preserving defect $D_N$~\eqref{eqn:h6dsym}, the
anomaly coefficients are
\begin{align}
  a = \frac{4000 \epsilon^2}{81 N^2} \left( 1 + \frac{186}{N} +
  \frac{60 492}{N^2} + \dots \right)\,,
  \qquad
  b_1 = -\frac{3b_2}{2} = -\frac{100 \epsilon}{27 N} \left( 1 + \frac{16}{N} -
  \frac{8172}{N^2} + \dots \right)\,.
  \label{eqn:anomalycoeff6dsym}
\end{align}
The coefficient $a$ is positive in accordance with the
$a$-theorem~\eqref{eqn:atheorem} if we interpret these defects as UV fixed
points of the dRG flow ending at the trivial defect.

\paragraph{Surface in $d = 4-\epsilon$.}
The previous calculation of the anomaly coefficients relies on the small distance
expansion of the 2-point function~\eqref{eqn:conformalprop}, which is known in
6d. Its generalisation to $d$ dimensions requires the 2-point function for an
operator $\cO^I$ of dimension 2 in a $d$ dimensional CFT. Requiring that it
reduces to the usual 2-point function when the metric is flat sets 
\begin{align}
  \vev{\cO^I(0) \cO^J(\xi)}_g
  =
  \frac{C_\cO \delta^{IJ}}{|\xi|^{2\Delta_\cO}}\left[ 1 + (\alpha R_{\mu\nu} +
    \beta R g_{\mu\nu}) \xi^{\mu} \xi^\nu + \dots \right]\,,
  \label{eqn:2ptsOcurvedspace}
\end{align}
with $\alpha, \beta$ constants that could a priori depend on $d$ and
$\Delta_\cO$. One can check that the UV divergences calculated as
in~\cite{Henningson:1999xi,Gustavsson:2004gj,Drukker:2020dcz} respect the
structure of a conformal anomaly~\eqref{eqn:anomalycoeff} only when
$R_{\mu\nu}$ and $R$ assemble into the Schouten tensor $P_{\mu\nu}$ with
coefficient $1/3$. Therefore we conclude that by consistency, the correlator
must take the same form as the propagator for scalars in
6d~\eqref{eqn:conformalprop} and the previous calculation extends to any
dimension $d$.

For the symmetry preserving defect~\eqref{eqn:hn4d}, we use~\eqref{eqn:anomalya}
and~\eqref{eqn:anomalycoeffres} to get
\begin{align}
  a = -\frac{9 N \epsilon^3}{2(N+8)^3}\,, \qquad
  b_1 = -\frac{9 N \epsilon^2}{4(N+8)^2}\,, \qquad
  b_2 = \frac{3 N \epsilon^2}{2(N+8)^2}\,.
  \label{eqn:anomalycoeffsres4dsym}
\end{align}

For the symmetry breaking defect~\eqref{eqn:4dsurfacebreakingsol}, there is a
contribution to the anomaly from the symmetry breaking term
$(\partial n)^2 = \partial_a n^{ij} \partial_b n^{ij} h^{ab}$, where
$h_{ab}$ is the induced metric on $\Sigma$ and $n^{ij}$ is the unit norm tensor
parallel to $h^{ij}$. The anomaly coefficients $b_1, b_2, c$ are straightforward
to obtain and given by
\begin{align}
  b_1 = -\frac{c}{2}\,, \qquad
  b_2 = \frac{c}{3}\,, \qquad
\end{align}
along with
\begin{equation}
\begin{aligned}
  c = \frac{h^{ij} h^{ij}}{64\pi^2}
  &=
  \frac{p h_{p,+}^2 + (N-p) h_{N-p,-}^2}{16\pi^2}\\
  &=
  \frac{3 \epsilon^2}{8 (N+8)^2} \left( 3 N + 2p(N-p) + (N-2p) \Delta \right)\,.
\end{aligned}
\end{equation}
Note that when $p = N$, $c \neq 0$ but $(\partial n)^2 = 0$, so we recover the
result~\eqref{eqn:anomalycoeffsres4dsym}.

For the coefficient $a$, we need to treat separately each irreducible
representation to get
\begin{align}
  a =
  \frac{\pi^2}{6} (\Delta_{\varphi^2_S}-2) C_{\varphi^2_S}
  \frac{(h_{ii})^2}{4}
  + \frac{\pi^2}{6} (\Delta_{\varphi^2_T}-2) (C_{\varphi^2_T})_{ijkl}
  \frac{h^{ij} h^{kl}}{4}\,.
\end{align}
The coefficient $C_{\varphi^2_T}$ can be read from the 2-point function of
$\varphi^2_T$. Plugging in the conformal dimensions along with the values of
$h$, we find
\begin{align}
  a
  =
  \frac{\epsilon^3}{48(N+8)^3}
  \left[ 8 \Delta^4 + 4 (N-2p) \Delta^3 + (N+6)(N-6) \Delta^2 - 9 (N+6)^2 \right]\,.
  \label{eqn:anomalycoeffap4d}
\end{align}
For any $N \le 6$ and $0 \le p \le N$, we can check that $\Delta > 0$ and $a$ satisfies
\begin{align}
  a_{p=0} \ge a_{p=1} \ge \dots \ge a_{p=N}\,.
\end{align}
This has a direct implication for defect RG flows. For any given $N$, one can
consider dRG
flows between symmetry breaking surface defects $D_{p}$ and $D_{p'}$. Because of the
$a$-theorem $a_{UV} > a_{IR}$~\cite{Jensen:2015swa}, we conclude that dRG flows
can only increase $p$, so that $p' > p$. We then expect a sequence of dRG flows that interpolate
between defects of increasing $p$
\begin{align}
  D_0 \to D_{1} \to \dots \to D_N\,,
\end{align}
with a concrete example being the flow $D_0 \to D_1 \to D_2$ presented in
figure~\ref{fig:rgflowex}. It would be interesting to check if this structure is
preserved at higher orders in the $\epsilon$-expansion.

\paragraph{Large $N$.}
Finally we can calculate the anomaly coefficients at large $N$ as well.
Using~\eqref{eqn:anomalya} and~\eqref{eqn:anomalycoeffres} we get
\begin{align}
  a = \frac{C_\sigma^3 \gamma_\sigma^3}{6C_{\sigma\sigma\sigma}^2}\,, \qquad
  b_1 = -\frac{C_\sigma^3 \gamma_\sigma^2}{2 C_{\sigma\sigma\sigma}^2}\,, \qquad
  b_2 = \frac{C_\sigma^3 \gamma_\sigma^2}{3 C_{\sigma\sigma\sigma}^2}\,, \qquad
  c = 0\,,
\end{align}
where $C_{\sigma\sigma\sigma}$ is given in~\eqref{eqn:largenCsss},
$\gamma_\sigma$ in~\eqref{eqn:gammas} and $C_\sigma$ in~\eqref{eqn:2ptsfuncs}.
Plugging in the values we get
\begin{align}
  a
  =
  \frac{1}{N^2} \frac{(d-2)(d-1) d^2}{3 (d-3)^2}
  \frac{\Gamma(d)^2\,\mathrm{sinc}\left( \frac{\pi d}{2} \right)}
  {\Gamma\left( \frac{d-4}{2} \right) \Gamma\left( \frac{d+2}{2} \right)^3}\,,
\end{align}
and
\begin{align}
  \frac{C_\sigma^3 \gamma_\sigma^2}{C_{\sigma\sigma\sigma}^2}
  =
  \frac{1}{N} \frac{(d-4)(d-2)(d-1)}{2(d-3)^2}
  \frac{\Gamma(d)\,\mathrm{sinc}\left( \frac{\pi d}{2} \right)}{\Gamma\left(
    \frac{d+2}{2} \right)}\,.
\end{align}
Setting $d = 4-\epsilon$ and $d = 6 - \epsilon$ we recover the large $N$ limit
of the anomaly coefficients~\eqref{eqn:anomalycoeffsres4dsym}
and~\eqref{eqn:anomalycoeff6dsym} as expected. Also notice that $a < 0$ when
$2 < d < 4$ and $a > 0$ when $4 < d < 6$, in accordance with the $a$-theorem.

\subsection{The string potential}

Given two planar surface defects separated a distance $L$, it is well-known that
quantum fluctuations can generate a potential $\frac{U_0 \mathrm{Area}}{L^2}$. A
generalisation of this observable called the (generalised) string potential was
introduced in~\cite{Drukker:2022beq} and is related to the anomaly coefficient
$b_1$, at least to leading order in the small perturbative parameter ($\epsilon$
or $N^{-1}$).

Consider the surface consisting of two hemispheres inside an
$S^3$ of radius $R$ and glued along their boundary at an angle $\pi -\phi$. Using
$u \in [0,\pi/2]$, $v \in [0, 2\pi)$ each hemisphere can be parametrised by
\begin{equation}
\begin{aligned}
  x^1 &= R \cos u \cos v \,, \qquad
  x^2 = R \cos u \sin v\,, \qquad\\
  x^3 &= R \sin u  \cos w\,, \qquad
  x^4 = R \sin u \sin w\,,
\end{aligned}
  \label{eqn:screaseparam}
\end{equation}
for a fixed $w$, and we take respectively $w = \phi/2$ and $w = \pi - \phi/2$
for each hemisphere.
When $\phi = 0$, this is the geometry of a sphere, while in the limit $\phi \to
\pi$ the two hemispheres become superimposed.
As shown in~\cite{Drukker:2022beq} the expectation value of a surface defect
$D_\phi$ with this geometry defines a potential $U(\phi)$ through
\begin{align}
U(\phi)=\frac{1}{2\pi}\big( \log{\vev{D_{\phi}}} -
\log{\vev{D_{\phi=0}}}\big)\,.
  \label{eqn:potentialdefn}
\end{align}
This definition is easy to motivate. The defect $D_\phi$ is topologically a
sphere, so it has a conformal anomaly and its expectation value is ill-defined.
However the anomaly cancels exactly in the ratio between the expectation values
at any two angles (here $\phi$ and 0), leading to a well-defined quantity.
The factor $2\pi$ is simply the area of the hemisphere, so that $U$ calculates
a potential density.

The expectation value of $D_\phi$ can be calculated in terms of an integrated
2-point function of the form~\eqref{eqn:intpropanomaly} and yields, to leading
order~\cite{Drukker:2022beq},
\begin{align}
  U(\phi)
  =
  \frac{b_1}{2\pi \cos(\phi/2)^2} + \ldots
\end{align}
In particular, setting $\phi = \pi - \delta$ and taking the limit $\delta \to 0$
we get
\begin{align}
  U(\pi - \delta) =
  \frac{2b_1}{\pi \delta^2} + \ldots
\end{align}
This can be interpreted as $\frac{U_0}{L^2}$ with the identification $L =
\delta$.
Note that $U_0$ is negative, so the force between the defects is attractive, as
expected for a force mediated by a scalar field.
Up to a different value for
$h_*$, this agrees with the result of~\cite{Rodriguez-Gomez:2022gbz}.

It would be interesting to determine whether the relation to the anomaly
coefficients still holds at subleading orders.

\section{Discussion}

Surface defects are ubiquitous objects in CFTs, and have been mostly studied in
the context of gauge theories where they play a fundamental role (e.g. in 4d
$\cN = 4$ SYM~\cite{Gukov:2006jk} and in the 6d $\cN = (2,0)$
theories~\cite{Gustavsson:2004gj,Drukker:2020dcz}).  The present analysis 
shows that even in the simplest example of an interacting CFT there is an
interesting array of surface defects. Thanks to this simplicity, one can 
perform perturbative calculations in overlapping regimes of validity, track dRG
flows, study their fixed points and provide new nontrivial examples of the
defect $a$-theorem.

The results obtained here suggest an interesting structure of fixed points
across dimensions and upon varying $N$. The simplest defect to access is the
symmetry preserving defect $D_N$, which seems to exist for $d \ge 3$ and any
$N$.  The most interesting case is $d = 3$, where the perturbative analysis
breaks down and the coupling to $\varphi^2$ changes scaling with respect to $N$.
I interpret this as evidence that $D_N$ matches with the fixed point of the
ordinary transition, but I haven't proven it.  It would be interesting to
confirm this expectation by extending the large $N$ methods
of~\cite{Giombi:2020rmc} to subleading order in $N$ and to surface defects in
arbitrary dimensions.

Two related questions that I haven't attempted to address here are diagnosing
the nature of two instabilities in the dRG flows. When $d < 4$, the trivial
defect is unstable. In this paper I discuss flows with $h > 0$ ending at
$D_N$, but as apparent in figure~\ref{fig:rgflowex}, dRG flows with $h < 0$ seem
to belong to a different universality class, and interpreting $h < 0 $ as a
negative mass term, we expect spontaneous symmetry breaking leading to the
extraordinary fixed point (i.e. where $\vev{\varphi^i} \neq 0$). When $d > 4$,
the direction of the flow is reversed and the defect $D_N$ becomes unstable.
Turning on the defect operator $\hat{\sigma}$ with one sign leads to the trivial
defect and is discussed in sections~\ref{sec:largen} and~\ref{sec:pert6d}. With
the other sign, it flows towards $h \to \infty$. That instability may be related
to the unbounded potential for $\sigma$ in~\eqref{eqn:S6d}.

In addition to $D_N$, the theory contains many instances of symmetry breaking
defects. At $d = 6 - \epsilon$ I find a (nonunitary) fixed
point~\eqref{eqn:hu6dsymbreak} that may correspond
to the extraordinary fixed point. Away from $\epsilon \ll 1$ the fixed point
cannot be studied reliably using the present methods, but is expected to exist for
any $2 < d < 6$ and any $N$. It would be interesting to study this fixed point
using large $N$ methods, and clarify whether there are additional
nonperturbative fixed points with $h, u \sim \sqrt{N}$ as suggested by the
perturbative beta functions~\eqref{eqn:beta6d}.

As one lowers $d$, more defect operators become relevant and the structure of the fixed
points becomes more intricate. Below $d = 4$, one may include a coupling to
$\varphi^i \varphi^j$, which leads to symmetry breaking defects preserving
$O(p) \times O(N-p) \subset O(N)$. Their behavior is dictated by the sign of the
discriminant $\Delta^2$~\eqref{eqn:4dsurfacebreakingsol}. It would be
interesting to understand the behavior of $\Delta$ as we vary $d$, especially
whether these fixed points still exist for some $N \le N_c$ at $d = 3$.
More generally, I expect that a full characterisation of fixed points near
$d=3$ would involve also the coupling to $\varphi^i$ as well as cubic and
quartic terms in $\varphi$, and it would be interesting to include these terms
in the analysis.

Although the focus of this paper is the critical $O(N)$ model, many more
examples of vector models where the $O(N)$ symmetry is reduced to a subgroup
were constructed in the $\epsilon$
expansion~\cite{Pelissetto:2000ek,Osborn:2017ucf,Rychkov:2018vya}.
A systematic study of surface defects in these models (as well as
generalisations involving complex scalars and fermions) as initiated
in~\cite{Pannell:2023pwz} for line defects is within reach and would expose
an even richer spectrum of surface defects.

Finally, the 3d (5d) critical $O(N)$ model is expected to be dual to a higher spin
theory on $AdS_4$~\cite{Klebanov:2002ja} ($AdS_6$), and it would be interesting
to study the realisation of surface operators there.

\subsection*{Acknowledgements}

It is a pleasure to thank Nadav Drukker, Zohar Komargodski, Diego
Rodriguez-Gomez, Ritam Sihna, Andy Stergiou, Volodia Schaub and Yifan Wang for
stimulating discussions. Special thanks go to N.  Drukker and A. Stergiou for
their helpful guidance, thoughtful comments on the preliminary version of this
manuscript and many suggestions. I also want to thank M. Probst and D.
Rodriguez-Gomez for sharing their notes on related topics with me.

\appendix

\section{Explicit calculations}
\label{app:pert4d}

This appendix presents the derivation of the beta functions governing the
renormalisation of the defect couplings for the surface operators introduced
in~\eqref{eqn:defnsurface4d} and~\eqref{eqn:surface6d}. I calculate the
relevant diagrams, with the results tabulated in tables~\ref{tab:feyn4d},
\ref{tab:feyn6dh} and \ref{tab:feyn6dn}, from which I read the corresponding
counterterms. The derivation of the beta function in $d = 4-\epsilon$ is
presented in section~\ref{sec:pert4d}, and in section~\ref{app:pert6d} I
present a more explicit derivation of the beta function in $d = 6 - \epsilon$.

\subsection{Evaluating Feynman diagrams}

Many of the Feynman diagrams we need are reducible to products of 1-loop
diagrams, and can be evaluated easily in momentum space
by using the identity (see e.g.~\cite{kleinert2001critical})
\begin{align}
  \int \frac{\diff^d k}{(2\pi)^d}
  \frac{1}{(k^2)^a [(k-p)^2]^b}
  =
  \frac{1}{(p^2)^{a+b-d/2}} L_d^{(a,b)}\,,
  \label{eqn:1loopintid}
\end{align}
where we defined
\begin{align}
  L_d^{(a,b)} =
  \frac{\Gamma\left( \frac{d}{2}-a \right)\Gamma\left( \frac{d}{2}-b
  \right)\Gamma\left( a+b- \frac{d}{2}
\right)}{(4\pi)^{d/2} \Gamma(a)\Gamma(b)\Gamma(d-a-b)}\,.
\label{eqn:defnl}
\end{align}
To revert back to position space we use the Fourier transform
\begin{align}
  \int \frac{\diff^d p}{(2\pi)^d} \frac{e^{ipx}}{(p^2)^{\alpha}}
  =
  C_{d,\alpha}
  \frac{1}{(x^2)^{\frac{d}{2} - \alpha}}\,,
  \qquad
  C_{d, \alpha}
  \equiv
  \frac{\Gamma \left( \frac{d}{2} - \alpha \right)}{4^\alpha \pi^{d/2}
  \Gamma(\alpha)}\,.
  \label{eqn:fourier}
\end{align}

As an example, we evaluate the diagram~\ref{fig:renormalisationh4dl0a}. Up
to a prefactor, this is given by the integral of (two copies of) the propagator
$G(x;y)$
\begin{align}
  M^{2 \epsilon}
  \int \diff^2 \tau G(x;\tau)^2\,,
\end{align}
with the factor $M^{2\epsilon}$ ensuring that the integral is dimensionless.
The momentum space representation of the propagator is given as usual by (this
is the $\alpha = 1$ case of~\eqref{eqn:fourier})
\begin{align}
  G(x;y)
  \equiv
  \frac{C_{d,1}}{|x-y|^{d-2}}
  =
  \int \frac{\diff^d p}{(2\pi)^d} \frac{e^{i p (x-y)}}{p^2}\,.
  \label{eqn:onpropagators}
\end{align}
Performing the $\tau$ integral and using~\eqref{eqn:1loopintid} leaves us with
\begin{align}
  M^{2 \epsilon}
  \int \frac{\diff^{d-2} p}{(2\pi)^{d-2}} \frac{ \diff^d p'}{(2\pi)^d}
  \frac{e^{i p x}}{p'^2 (p-p')^2}
  &=
  M^{2 \epsilon}
  \int \frac{\diff^{2-\epsilon} p}{(2\pi)^{2-\epsilon}}
  \frac{e^{i p x}}{p^{\epsilon}}
  L_{4-\epsilon}^{(1,1)}
  =
  \frac{(x M)^{2 \epsilon}}{x^2} C_{2-\epsilon,\frac{\epsilon}{2}}
  L_{4-\epsilon}^{(1,1)}\,.
\end{align}

Using these identities we can evaluate most of the diagrams entering the
perturbative calculations in 4d, 6d and at large $N$. One exception is the
diagram~\ref{eqn:feyn4dh2}, which we analyse below.

\begin{table}[h]
  \centering
  \renewcommand{\arraystretch}{1.5}
  \begin{tabular}{crll}
    \toprule
    \midrule
  Diagram & Prefactor & Integral & $\epsilon$-expansion\\
  \midrule
  \FDfa &
  $-h_{ij}$ &
  $\frac{(x M)^{2\epsilon}}{x^2}
  C_{2-\epsilon,\frac{\epsilon}{2}}\,L_{4-\epsilon}^{(1,1)}$
  & $\frac{1}{16 \pi^3 x^2} + \dots$ \\
  \FDfb &
  $h_{ik} h_{kj}$
  & $ \frac{(xM)^{3 \epsilon}}{x^2}
  C_{4-\epsilon,1} T_1 $
  & $\frac{1}{32 \pi^4 x^2 \epsilon} + \dots$ \\
  \FDfc
  & $\frac{\lambda}{2} \frac{h_{KK} \delta_{ij} + 2 h_{ij}}{3}$ &
  $\frac{(x M)^{3\epsilon}}{x^2}
  C_{2-\epsilon,\epsilon}\,(L_{4-\epsilon}^{(1,1)})^2$
  & $\frac{1}{64 \pi^5 x^2 \epsilon} + \dots$\\
  \hline
  \FDfca
  & $-\delta h_{ij}$
  & $\frac{(x M)^{2\epsilon}}{x^2}
  C_{2-\epsilon,\frac{\epsilon}{2}}\,L_{4-\epsilon}^{(1,1)}$
  & $\frac{1}{16 \pi^3 x^2} + \dots$ \\
  \FDfcb
  & $\delta_{\varphi^2_R} h_{ij}$
  &
  $\frac{(x M)^{2\epsilon}}{x^2}
  C_{2-\epsilon,\frac{\epsilon}{2}}\,L_{4-\epsilon}^{(1,1)}$
  & $\frac{1}{16 \pi^3 x^2} + \dots$ \\
  \bottomrule
  \end{tabular}
  \renewcommand{\arraystretch}{1}
  \caption{Feynman diagrams contributing to the renormalisation of $h_{ij}$ to
order $O(\lambda^2, \lambda h^2, h^3)$. We list the value of the integral
of the propagators, its leading pole in the $\epsilon$-expansion, and the
overall prefactor of the integral. The values for $C_{d,\alpha}$ and
$L_{d}^{(a,b)}$ are respectively given in~\eqref{eqn:fourier}
and~\eqref{eqn:defnl}. The counterterm $\delta_{\varphi^2_R}$ is defined
in~\eqref{eqn:deltaphi2}.}
  \label{tab:feyn4d}
\end{table}

\paragraph{Diagram~\ref{eqn:feyn4dh2}.} It is useful to consider the more
general integral
\begin{align}
  M^{(m+2)\epsilon}
  \int \diff^2 \tau_1 \diff^2 \tau_2
  \frac{ C_{d,1}^3}{|x - \tau_1|^{2 - \epsilon}|\tau_1 - \tau_2|^{2 - m\epsilon}|x -
  \tau_2|^{2 - \epsilon}}\,.
  \label{eqn:defn0T}
\end{align}
When $m=1$, this reduces to the diagram~\ref{eqn:feyn4dh2}.
Using either momentum of position space propagators, one can reduce this to the
integral over Feynman parameters
\begin{align}
  \frac{(xM)^{(m+2) \epsilon}}{x^2}
  C_{d,1}
  T_m\,,
\end{align}
where we defined
\begin{align}
  T_m
  =
  \frac{\Gamma\left( 1 - \frac{(m+2) \epsilon}{2} \right)}{16\pi^{2-\epsilon}
  \Gamma\left( 1 - \frac{m \epsilon}{2} \right)}
  \int\limits_0^1 \diff u \int\limits_0^1 \diff v
  u^{-1 + \frac{m\epsilon}{2}} v^{-\frac{\epsilon}{2}}
  [(1 - u)(1-v)]^{-\frac{(m+1)\epsilon}{2}}
  (1-u v)^{-1 + \frac{(m+2) \epsilon}{2}}\,.
\end{align}
This integral is easy to do by expanding $(1-uv)^{-1+\frac{(m+2)\epsilon}{2}}$ in
series, in which case the integral factorises into two beta functions to give
\begin{align}
  \frac{\Gamma\left( 1 - \frac{(m+2) \epsilon}{2} \right)
  \Gamma\left( 1 - \frac{(m+1) \epsilon}{2} \right)^2
  \Gamma\left( \frac{(m+2)\epsilon}{2} \right)}
  {16\pi^{2-\epsilon}\Gamma\left( 1 - \frac{m \epsilon}{2} \right)}
  \sum_{n \ge 0}
  \frac{\Gamma\left( n + \frac{m\epsilon}{2} \right)}
  {\Gamma\left( n+2-\frac{(m+2)\epsilon}{2} \right)
  \Gamma\left( \frac{(m+2)\epsilon}{2} - n \right)}
  \frac{(-1)^n}{n!}\,.
\end{align}
The sum can be written as a hypergeometric function and evaluated to find
\begin{align}
  T_m
  =
  \frac{
    \Gamma\left( 1-\frac{(m+1)\epsilon}{2} \right)^2
  \Gamma\left( 1 - \frac{(m+2)\epsilon}{2} \right)
  \Gamma\left( \frac{m\epsilon}{2} \right)}
  {16 \pi^{2-\epsilon} \Gamma\left( 2 - (m+1) \epsilon \right)}\,.
  \label{eqn:defnT}
\end{align}

\subsection{An all-loop result in $d = 4 - \epsilon$}
\label{sec:allloop}

When the coupling $h$ is small, its beta function is given
by~\eqref{eqn:beta4d}. In this appendix I evaluate the diagrams relevant for
determining the beta function at order $h \sim O(\epsilon^0)$ exactly. The
leading contribution is the beta function for surface defects of the free
$O(N)$ model, and we find that there are no new fixed points of order $h_* \sim
O(\epsilon^0)$, in agreement with the bootstrap analysis
of~\cite{Lauria:2020emq}.

Assuming $h \sim \epsilon^0$, the leading diagrams that contribute to the
renormalisation of the defect coupling $h$ are
\begin{align}
  \FDfa + \FDfb + \FDfba + \dots
  \label{eqn:4dalldiag}
\end{align}
Introducing the notation
\begin{align}
  \FDfbb
  =
  ( M^\epsilon h_{ij})^m
  \int \diff^2 \tau_i \diff^2 \tau_{i+1}
  \frac{C_{d,1}}{|\tau_i - \tau_{i+1}|^{2-m \epsilon}}\,,
\end{align}
we have the recursion relation
\begin{align}
  \FDfbc
  =
  \frac{C_{d,1} \pi \Gamma\left( \frac{\epsilon}{2} \right) \Gamma\left(
  \frac{m \epsilon}{2} \right) \Gamma\left( 1 - \frac{(m+1) \epsilon}{2} \right)}
  {\Gamma\left( 1 - \frac{\epsilon}{2} \right) \Gamma\left(
  \frac{(m+1)\epsilon}{2} \right) \Gamma\left(1 - \frac{m \epsilon}{2} \right)}
  \times \FDfbd\,.
\end{align}
The diagram with $m$ legs ending on the defect can then be reduced to the integral
\begin{align}
  (h_{ij})^m
  \frac{(\pi C_{d,1})^{m-1} \Gamma\left( \frac{\epsilon}{2} \right)^m
  \Gamma\left(1 - \frac{m \epsilon}{2}\right)}
  {\Gamma\left( 1 - \frac{\epsilon}{2} \right)^m \Gamma\left( \frac{m \epsilon}{2} \right)}
  \int \diff^2 \tau_1 \diff^2 \tau_m
  \frac{ M^{(m+2)\epsilon} C_{d,1}^3}{|x - \tau_1|^{2 - \epsilon}|\tau_1 - \tau_n|^{2 - m\epsilon}|x -
  \tau_m|^{2 - \epsilon}}\,.
\end{align}
This last integral is $T_m$~\eqref{eqn:defn0T}, with the result given in~\eqref{eqn:defnT}.

Summing all diagrams~\eqref{eqn:4dalldiag} up to order $h^n$, we can absorb all
the poles in $\epsilon$ by taking the counterterm to be
\begin{align}
  \delta h
  =
  h \sum_{m = 1}^{n-1} \left( \frac{h}{2 \pi \epsilon} \right)^m
  + O(\lambda)\,.
\end{align}
Taking the limit $n \to \infty$ and calculating the beta function, we find
\begin{align}
  \beta_{h_{ij}} =
  - \epsilon h_{ij} + \frac{1}{2\pi}h_{ij} h_{ji} + O(\lambda)\,.
\end{align}
This is an exact result in $h$ and it agrees with~\eqref{eqn:beta4d}. Since
$\lambda_* \sim O(\epsilon)$ and the beta function must vanish order by order in
$\epsilon$ at the fixed points, we conclude that also in the interacting theory
there are no fixed point of order $O(\epsilon^0)$.

\subsection{Renormalisation and defect beta function in 6d}
\label{app:pert6d}

\begin{table}[h]
  \centering
  \begin{tabular}{crll}
  \toprule \midrule
  Diagrams & Prefactor & Integral & $\epsilon$-expansion\\
  \midrule
  \FDsa
  & $-h$
  & $\frac{(x M)^{\epsilon}}{x^2} C_{4-\epsilon,1}$
  & $\frac{1}{4\pi^2 x^2} + \dots$ \\
  \FDsb $+$ \FDsf &
  $-\frac{h^2 g_2 + u^2 g_1}{2}$
  & $\frac{(x M)^{2\epsilon}}{x^2} C_{4-\epsilon,1+\frac{\epsilon}{2}}
  L_{4-\epsilon}^{(1,1)}$
  & $\frac{1}{32 \pi^4 x^2 \epsilon} + \dots$\\
  \FDsc $+$ \FDsd
  & $-\frac{h (N g_1^2 + g_2^2)}{2}$
  & $\frac{(x M)^{2\epsilon}}{x^2} C_{4-\epsilon,1+\frac{\epsilon}{2}}
  L_{6-\epsilon}^{(1,1)}$
  & $\frac{-1}{768 \pi^5 x^2 \epsilon} + \dots$\\
  \FDse $+$ \FDsg
  & \multirow{2}{*}{$-\frac{h^3 g_2^2 + h u^2 g_1 (g_2 + 2 g_1)}{2}$}
  & \multirow{2}{*}{$\frac{(x M)^{3\epsilon}}{x^2} C_{4-\epsilon,1+\epsilon}
  L_{4-\epsilon}^{(1,1)}L_{4-\epsilon}^{(1,1+\frac{\epsilon}{2})}$}
  & \multirow{2}{*}{$\frac{1}{512\pi^6 x^2 \epsilon^2} + \dots$}\\
  $+$ \FDsh & & &\\
  \midrule
  \FDsca
  & $-\delta h$
  & $\frac{(x M)^{\epsilon}}{x^2} C_{4-\epsilon,1}$
  & $\frac{1}{4\pi^2 x^2} + \dots$ \\
  \FDscb
  & $h \delta_\sigma$
  & $\frac{(x M)^{\epsilon}}{x^2} C_{4-\epsilon,1}$
  & $\frac{1}{4\pi^2 x^2} + \dots$ \\
  \FDscc
  & $-h g_2 \delta h$
  & $\frac{(x M)^{2\epsilon}}{x^2} C_{4-\epsilon,1+\frac{\epsilon}{2}}
  L_{4-\epsilon}^{(1,1)}$
  & $\frac{1}{32 \pi^4 x^2 \epsilon} + \dots$\\
  \bottomrule
  \end{tabular}
  \renewcommand{\arraystretch}{1}
  \caption{Feynman diagrams contributing to the renormalisation of $h$ to
order $O(g^2)$. We list the value of the integral
of the propagators, its leading pole in the $\epsilon$-expansion, and the
overall prefactor of the integral. The values for $C_{d,\alpha}$ and
$L_{d}^{(a,b)}$ are respectively given in~\eqref{eqn:fourier}
and~\eqref{eqn:defnl}.}
  \label{tab:feyn6dh}
\end{table}

\begin{table}[h]
  \centering
  \begin{tabular}{crll}
  \toprule \midrule
  Diagrams & Prefactor & Integral & $\epsilon$-expansion\\
  \midrule
  \FDsj
  & $-u^i$
  & $\frac{(x M)^{\epsilon}}{x^2} C_{4-\epsilon,1}$
  & $\frac{1}{4\pi^2 x^2} + \dots$ \\
  \FDsk &
  $- u^i h g_1$
  & $\frac{(x M)^{2\epsilon}}{x^2} C_{4-\epsilon,1+\frac{\epsilon}{2}}
  L_{4-\epsilon}^{(1,1)}$
  & $\frac{1}{32 \pi^4 x^2 \epsilon} + \dots$\\
  \FDsp
  & $-u^i g_1^2$
  & $\frac{(x M)^{2\epsilon}}{x^2} C_{4-\epsilon,1+\frac{\epsilon}{2}}
  L_{6-\epsilon}^{(1,1)}$
  & $\frac{-1}{768 \pi^5 x^2 \epsilon} + \dots$\\
  \FDsm $+\FDsn$
  & \multirow{2}{*}{$-\frac{u^i \left( u^2 g_1^2 + h^2 g_1 (g_2 + 2 g_1)
  \right)}{2}$}
  & \multirow{2}{*}{$\frac{(x M)^{3\epsilon}}{x^2} C_{4-\epsilon,1+\epsilon}
  L_{4-\epsilon}^{(1,1)}L_{4-\epsilon}^{(1,1+\frac{\epsilon}{2})}$}
  & \multirow{2}{*}{$\frac{1}{512\pi^6 x^2 \epsilon^2} + \dots$}\\
  \FDsl & & &\\
  \midrule
  \FDsca
  & $-\delta u^i$
  & $\frac{(x M)^{\epsilon}}{x^2} C_{4-\epsilon,1}$
  & $\frac{1}{4\pi^2 x^2} + \dots$ \\
  \FDscb
  & $u^i \delta_\varphi$
  & $\frac{(x M)^{\epsilon}}{x^2} C_{4-\epsilon,1}$
  & $\frac{1}{4\pi^2 x^2} + \dots$ \\
  \FDscc
  & $- (\delta u^i h + u^i \delta h) g_1$
  & $\frac{(x M)^{2\epsilon}}{x^2} C_{4-\epsilon,1+\frac{\epsilon}{2}}
  L_{4-\epsilon}^{(1,1)}$
  & $\frac{1}{32 \pi^4 x^2 \epsilon} + \dots$\\
  \bottomrule
  \end{tabular}
  \renewcommand{\arraystretch}{1}
  \caption{Feynman diagrams contributing to the renormalisation of $u^i$ to
order $O(g^2)$. We list the value of the integral
of the propagators, its leading pole in the $\epsilon$-expansion, and the
overall prefactor of the integral. The values for $C_{d,\alpha}$ and
$L_{d}^{(a,b)}$ are respectively given in~\eqref{eqn:fourier}
and~\eqref{eqn:defnl}.}
  \label{tab:feyn6dn}
\end{table}

In this section I present the calculation of the beta function for the surface
defects~\eqref{eqn:surface6d}.
The calculation is performed in the minimal subtraction (MS) scheme, where the
counterterms $\delta h, \delta u^i$ absorb poles in $\epsilon$ and admit the expansion
\begin{align}
  \delta h
  =
  \frac{\delta h^{(1)}}{\epsilon} + \frac{\delta h^{(2)}}{\epsilon^2}
  + \dots
  \qquad
  \delta u^i
  =
  \frac{(\delta u^i)^{(1)}}{\epsilon} + \frac{(\delta u^i)^{(2)}}{\epsilon^2} +
  \dots
\end{align}
The counterterms for the bulk theory have been calculated previously, and here
we need~\cite{Fei:2014yja}
\begin{align}
  \delta_\sigma 
  =
  -\frac{N g_1^2 + g_2^2}{6 (4\pi)^3 \epsilon} + \dots\,,
  \qquad
  \delta_\varphi
  =
  -\frac{g_1^2}{3 (4\pi)^3 \epsilon} + \dots\,.
\end{align}

Requiring all poles in $\epsilon$ to cancel in the 1-point functions for $h$ and
$u^i$ fixes the counterterms, and we find
\begin{align}
  \delta h &=
  \frac{1}{\epsilon} \left[
  -\frac{g_1 u^2 + g_2 h^2}{(4\pi)^2}
  + \frac{(2g_1 + g_2) g_1 h u^2 + g_2^2 h^3}{2 (4\pi)^4} \right]
  - \frac{1}{\epsilon^2}
  \frac{(2 g_1 + g_2) g_1 h u^2 + g_2^2 h^3}{(4\pi)^4}
  + O(g^3)\,,\\
  \delta u^i &=
  \frac{u^i}{\epsilon} \left[ 
    -\frac{2 g_1 h}{(4\pi)^2}
    + \frac{(2g_1 + g_2) g_1 h^2 + g_1^2 u^2}{2(4\pi)^4}
  \right]
  - \frac{u^i}{\epsilon^2}
  \frac{(2g_1+g_2)g_1 h^2 + g_1^2 u^2}{(4\pi)^4}
  + O(g^3)\,.
\end{align}

From these we can read the beta function. The bare couplings are related to the
renormalised couplings via
\begin{align}
  h_0 = M^{\epsilon/2} (h + \delta h) Z_{\sigma}^{-1/2}\,,
  \qquad
  u^i_0 = M^{\epsilon/2} (u^i + \delta u^i) Z_{\varphi}^{-1/2}\,.
\end{align}
To obtain the beta function, we take $M \frac{d}{dM} \log h_0$ to get
\begin{align}
  0 = \frac{\epsilon}{2}
  + \left( \beta_h \frac{\partial}{\partial h}
  + \beta_{u^i} \frac{\partial}{\partial u^i}
    + \beta_{g_1} \frac{\partial}{\partial g_1}
    + \beta_{g_2} \frac{\partial}{\partial g_2}
  \right)
  \log\left( (h + \delta h) Z_\sigma^{-1/2}
  \right)\,,
\end{align}
and similarly for $u^i$.
Working at order $O(g^2)$, we have $\beta_g = -\frac{\epsilon g}{2} + \dots$, and we obtain
\begin{align}
  \beta_h = -\frac{\epsilon h}{2}
  + h \gamma_\sigma
  + \frac{h}{2} \partial_h \delta h^{(1)}
  + \frac{g_i}{2} \partial_{g_i} \delta h^{(1)}
  - \frac{1}{2} \delta h^{(1)}
  + O(g^3)\,.
\end{align}
A similar equation holds for $\beta_n$. Plugging the value for the counterterms
we get~\eqref{eqn:beta6d}.

\bibliographystyle{utphys2}
\bibliography{ref}

\end{document}